\documentclass[%
 reprint,
 superscriptaddress,
 amsmath,amssymb,
 aps,
 prb,
]{revtex4-2}

\setcitestyle{number}
\usepackage{graphicx}% Include figure files
\usepackage{amsmath,amsfonts,amssymb}
\usepackage{dcolumn}% Align table columns on decimal point
\usepackage{bm}% bold math
\usepackage{hyperref}% add hypertext capabilities
\hypersetup{
    colorlinks=true,
    linkcolor=blue,
    filecolor=magenta,      
    urlcolor=blue,
    citecolor=blue
    }
\begin{document}

%\preprint{APS/123-QED}

\title{Parallel Exploration of the Optoelectronic Properties of (Sb,Bi)(S,Se)(Br,I) Chalcohalides}% Force line breaks with \\

\author{Rasmus S. Nielsen}
\email[]{Electronic mail: rasmus.nielsen@empa.ch}
\affiliation{Nanomaterials Spectroscopy and Imaging, Transport at Nanoscale Interfaces Laboratory, Swiss Federal Laboratories for Material Science and Technology (EMPA), Ueberlandstrasse 129, 8600 Duebendorf, Switzerland}

\author{Ángel Labordet Álvarez}
\affiliation{Nanomaterials Spectroscopy and Imaging, Transport at Nanoscale Interfaces Laboratory, Swiss Federal Laboratories for Material Science and Technology (EMPA), Ueberlandstrasse 129, 8600 Duebendorf, Switzerland}
\affiliation{Department of Physics, University of Basel, 4056 Basel, Switzerland}
\affiliation{Swiss Nanoscience Institute, University of Basel, 4056 Basel, Switzerland}

\author{Axel G. Medaille}
\affiliation{Universitat Politècnica de Catalunya (UPC), Photovoltaic Lab - Micro and Nano Technologies Group (MNT), Electronic Engineering Department, EEBE, Av Eduard Maristany 10-14, Barcelona 08019, Spain}
\affiliation{Universitat Politècnica de Catalunya (UPC), Barcelona Centre for Multiscale Science \& Engineering, Av Eduard Maristany 10-14, Barcelona 08019, Spain}

\author{Ivan Caño}
\affiliation{Universitat Politècnica de Catalunya (UPC), Photovoltaic Lab - Micro and Nano Technologies Group (MNT), Electronic Engineering Department, EEBE, Av Eduard Maristany 10-14, Barcelona 08019, Spain}
\affiliation{Universitat Politècnica de Catalunya (UPC), Barcelona Centre for Multiscale Science \& Engineering, Av Eduard Maristany 10-14, Barcelona 08019, Spain}

\author{Alejandro Navarro-Güell}
\affiliation{Universitat Politècnica de Catalunya (UPC), Photovoltaic Lab - Micro and Nano Technologies Group (MNT), Electronic Engineering Department, EEBE, Av Eduard Maristany 10-14, Barcelona 08019, Spain}
\affiliation{Universitat Politècnica de Catalunya (UPC), Barcelona Centre for Multiscale Science \& Engineering, Av Eduard Maristany 10-14, Barcelona 08019, Spain}

\author{Cibrán L. Álvarez}
\affiliation{Group of Characterization of Materials, Departament de Física, Universitat Politècnica de Catalunya (UPC), Campus Diagonal-Besòs, Av Eduard Maristany 10–14, Barcelona 08019, Spain}
\affiliation{Universitat Politècnica de Catalunya (UPC), Barcelona Centre for Multiscale Science \& Engineering, Av Eduard Maristany 10-14, Barcelona 08019, Spain}

\author{Claudio Cazorla}
\affiliation{Group of Characterization of Materials, Departament de Física, Universitat Politècnica de Catalunya (UPC), Campus Diagonal-Besòs, Av Eduard Maristany 10–14, Barcelona 08019, Spain}
\affiliation{Universitat Politècnica de Catalunya (UPC), Barcelona Centre for Multiscale Science \& Engineering, Av Eduard Maristany 10-14, Barcelona 08019, Spain}

\author{David R. Ferrer}
\affiliation{Universitat Politècnica de Catalunya (UPC), Photovoltaic Lab - Micro and Nano Technologies Group (MNT), Electronic Engineering Department, EEBE, Av Eduard Maristany 10-14, Barcelona 08019, Spain}
\affiliation{Universitat Politècnica de Catalunya (UPC), Barcelona Centre for Multiscale Science \& Engineering, Av Eduard Maristany 10-14, Barcelona 08019, Spain}

\author{Zacharie J. Li-Kao}
\affiliation{Universitat Politècnica de Catalunya (UPC), Photovoltaic Lab - Micro and Nano Technologies Group (MNT), Electronic Engineering Department, EEBE, Av Eduard Maristany 10-14, Barcelona 08019, Spain}
\affiliation{Universitat Politècnica de Catalunya (UPC), Barcelona Centre for Multiscale Science \& Engineering, Av Eduard Maristany 10-14, Barcelona 08019, Spain}

\author{Edgardo Saucedo}
\affiliation{Universitat Politècnica de Catalunya (UPC), Photovoltaic Lab - Micro and Nano Technologies Group (MNT), Electronic Engineering Department, EEBE, Av Eduard Maristany 10-14, Barcelona 08019, Spain}
\affiliation{Universitat Politècnica de Catalunya (UPC), Barcelona Centre for Multiscale Science \& Engineering, Av Eduard Maristany 10-14, Barcelona 08019, Spain}

\author{Mirjana Dimitrievska}
\email[]{Electronic mail: mirjana.dimitrievska@empa.ch}
\affiliation{Nanomaterials Spectroscopy and Imaging, Transport at Nanoscale Interfaces Laboratory, Swiss Federal Laboratories for Material Science and Technology (EMPA), Ueberlandstrasse 129, 8600 Duebendorf, Switzerland}

\begin{abstract}

Chalcohalides are an emerging family of semiconductors with irresistible material properties, shaped by the intricate interplay between their unique structural chemistry and vibrational dynamics. Despite their promise for next-generation solar energy conversion devices, their intrinsic optoelectronic properties remain largely unexplored.  Here, we focus on the (Sb,Bi)(S,Se)(Br,I) system, a subset of compounds that share the same quasi-1D crystal structure. Using a two-step physical vapor deposition (PVD) process, we synthesize the eight ternary chalcohalide compounds, demonstrating bandgaps ranging from 1.38 to 2.08 eV with sharp, single-component photoluminescence (PL) peaks. In a parallel exploration of carrier dynamics and intrinsic electron–phonon interactions -- comprehensively studied using power-, temperature-dependent, and time-resolved PL measurements -- we map their direct impact on optoelectronic performance. Supported by first-principles density functional theory (DFT) defect calculations, we establish clear structure–property relations, identifying solid-solutions engineering as an effective means to fine-tune the native phonon structures and further suppress non-radiative recombination. This study provides a blueprint for optimizing chalcohalides as high-efficiency materials across a wide range of optoelectronic applications.

\end{abstract}

\maketitle

\section{Introduction}

Lead-halide perovskites have transformed the field of optoelectronics due to their remarkable material properties \cite{wang2021a, liu2021a, ren2022a, lei2021a, cheng2022a}, including high absorption coefficients \cite{de2014a}, long carrier diffusion lengths \cite{stranks2013a}, and an unusual tolerance to defects \cite{dequilettes2016a}. Despite these advantages, concerns about lead toxicity \cite{babayigit2016a, serrano2015a} and long-term instability \cite{duan2023a, conings2015a, leijtens2015a} have catalyzed the search for alternative materials. Chalcogenides, in contrast, have been explored as another promising class of optoelectronic materials, offering excellent chemical and thermal stability, but they generally fall short of matching the superior optoelectronic performance of halide perovskites \cite{kim2020a, Nielsen2025a}. The ultimate goal in developing next-generation optoelectronic technologies is to discover a new family of materials that combines the defect tolerance and electronic properties of halide perovskites with the exceptional environmental and operational stability of chalcogenides.

Chalcohalides are an emerging family of inorganic semiconductors that combine both chalcogen and halogen anions \cite{ghorpade2023a, zhang2025a, l2024a, guo2023a, palazon2022a, wla2018a}. The instability of halide perovskites is often attributed to the high electronegativity and ionic bonding character of halogen anions. In chalcohalides, however, the partial substitution of halogens with chalcogen anions -- sometimes referred to as a "split-anion approach" \cite{sun2016a} -- introduces a more covalent chemical bonding environment, which is predicted to enhance both chemical and thermal stability \cite{kavanagh2021a, liu2021b, roth2024a}. The coexistence of divalent chalcogen and monovalent halogen anions also enables a high degree of tunability in structural and optoelectronic properties. Moreover, the incorporation of trivalent metal cations offers an electronic configuration similar to Pb$^\text{2+}$, where the unique chemistry of the ns$^\text{2}$ lone-pair state is known to play a crucial role in the exceptional defect tolerance of lead-halide perovskites \cite{brandt2017a, kurchin2018a, nicolson2023a}. This suggests that chalcohalides, with their analogous electronic structure, may exhibit similar defect-tolerant properties. These distinctive electronic structures also feature highly dispersive band edges, leading to high carrier mobilities  \cite{ran2018a, ganose2016a}. Finally, by mitigating issues related to halogen volatility, chalcohalides are more thermodynamically stable than halide perovskites, further supporting their potential for integration into a wide range of optoelectronic devices.

Despite combining many of the most favorable material properties for optoelectronic applications, chalcohalides are still in the early stages of research. Unlocking their full potential and identifying the most promising candidates requires a deeper understanding of their fundamental properties. This includes not only advancing synthesis strategies to realize and optimize thin films, but also addressing the current lack of experimental studies on their optoelectronic properties. Moreover, given the vast compositional space accessible through different combinations of chalcogen and halogen anions with various metal cations, it would be more efficient to explore smaller subsets of chalcohalides -- such as those sharing similar crystal structures -- in parallel. Such targeted studies can accelerate the mapping of synthesis–structure–property relations, enabling precise functionality engineering \cite{zunger2018a}. These mapped relations may ultimately serve as a blueprint for the inverse design of next-generation optoelectronic materials with tailored properties.

In this work, we investigate the optoelectronic properties of a combinatorial subset of heavy pnictogen-based ternary chalcohalides with the general formula (Sb,Bi)(S,Se)(Br,I). These eight compounds all crystallize in the same quasi-one-dimensional orthorhombic structure with space group \textit{Pnma}, providing an ideal platform for comparative studies. We present a two-step synthesis route, in which chalcogenide precursors are first formed via thermal co-evaporation, followed by high-pressure reactive annealing to incorporate the halogen and form the final ternary phase. We first characterize their structural and morphological properties, alongside optical measurements to evaluate light absorption and emission behavior. Based on these results, we identify the most promising compounds for more in-depth photoluminescence (PL) studies, including power- and temperature-dependent measurements, as well as excitation intensity-dependent time-resolved PL spectroscopy (TRPL). These measurements are used to assess the nature of the emissive states, charge-carrier recombination dynamics, and electron-phonon interactions, offering insight into the fundamental processes that govern the intrinsic optoelectronic performance in this class of materials.

\section{Methods}

\small

Chalcohalide samples were synthesized using a physical vapor deposition process consisting of co-evaporation followed by reactive annealing. Initially, binary chalcogenides were deposited onto glass/Mo and quartz substrates via co-evaporation from elemental sources of Sb (Sigma-Aldrich, 100 mesh, 99.5\%), Bi (Sigma-Aldrich, 100 mesh, 99\%), Se (Thermo-Scientific, 200 mesh, 99.9\%), and S (Thermo-Scientific, flakes, 99.9\%). The evaporation temperatures were set to 550$^\circ$C for Sb and 680$^\circ$C for Bi, while the substrate temperature was maintained at 280$^\circ$C to facilitate immediate reaction with the chalcogen species.

Following deposition, the chalcogenide precursor was placed in a Petri dish along with the corresponding halide source -- SbI$_3$ (Thermo-Scientific, 99.999\%), SbBr$_3$ (Thermo-Scientific, 99.995\%), BiI$_3$ (Puratronic, 99.999\%), or BiBr$_3$ (Thermo-Scientific, 99\%). For example, Sb$_2$Se$_3$ was combined with SbI$_3$ to form SbSeI. The Petri dish was then sealed within a steel tubular furnace under an inert Ar atmosphere and subjected to high-temperature, high-pressure annealing. The annealing conditions were optimized for each chalcohalide composition, as detailed in Table \ref{tbl:AnnealingConditions}. All purities specified are metals basis. The synthesis process has been schematically illustrated in Figure S1. See Ref \cite{ca2025a} for further synthesis details.

\begin{table}[ht!]
\small
\caption{Annealing conditions used for the fabrication of chalcohalide compounds.}
\vspace{0.25cm}
\label{tbl:AnnealingConditions}
\begin{tabular*}{\columnwidth}{@{\extracolsep{\fill}}lccc} % Adjusted column alignment
    Compound & $T$ ($^\circ$C) & $p$ (bar) & t (min) \\ \hline
    BiSBr & 400 & 2.5 & 15 \\
    BiSI & 425 & 2.5 & 15 \\
    BiSeBr & 450 & 4.0 & 15 \\
    BiSeI & 500 & 4.0 & 15 \\
    SbSBr & 325 & 2.5 & 15 \\
    SbSI & 300 & 2.0 & 15 \\ 
    SbSeBr & 450 & 4.0 & 15 \\
    SbSeI & 450 & 3.0 & 15 \\
    \hline 
\end{tabular*}
\end{table}

Photoluminescence (PL) spectra were measured using a Horiba LabRam confocal microscope in a backscattering configuration with 538 nm and 488 nm excitation lasers. The excitation beam was focused onto the sample surface using a long-range 50× microscope objective with a numerical aperture of 0.55, yielding a beam diameter of approximately 1.1 $\mu$m for the 488 nm laser and 1.2 $\mu$m for the 532 nm laser. Temperature-dependent PL spectra were recorded from 70 K to 300 K using a MicroStat HiRes optical cryostat from Oxford Instruments with liquid nitrogen cooling, while excitation power-dependent PL studies were conducted at 70 K. The backscattered light was analyzed using a spectrometer equipped with a 150 g/mm holographic grating and a thermoelectrically cooled CCD.

Time-resolved photoluminescence (TRPL) measurements were conducted using a MicroTime 100 time-resolved confocal microscope coupled with a PicoQuant detection unit. Samples were excited using a pulsed 488 nm laser with a pulse duration of $<$100 ps and a repetition rate of 64 MHz. The excitation beam was focused onto the sample surface through a long-range 20x objective with a numerical aperture of 0.45. Emitted photons were collected through the same objective and guided via a 50 $\mu$m diameter optical fiber to the detection unit, which included a FluoTime 300 photospectrometer, a monochromator, and a photomultiplier detector. A cutoff filter was employed to suppress reflected laser photons. TRPL measurements were performed at room temperature.

\begin{figure*}[t!]
    \centering
    \includegraphics[width=\textwidth,trim={0 0 0 0},clip]{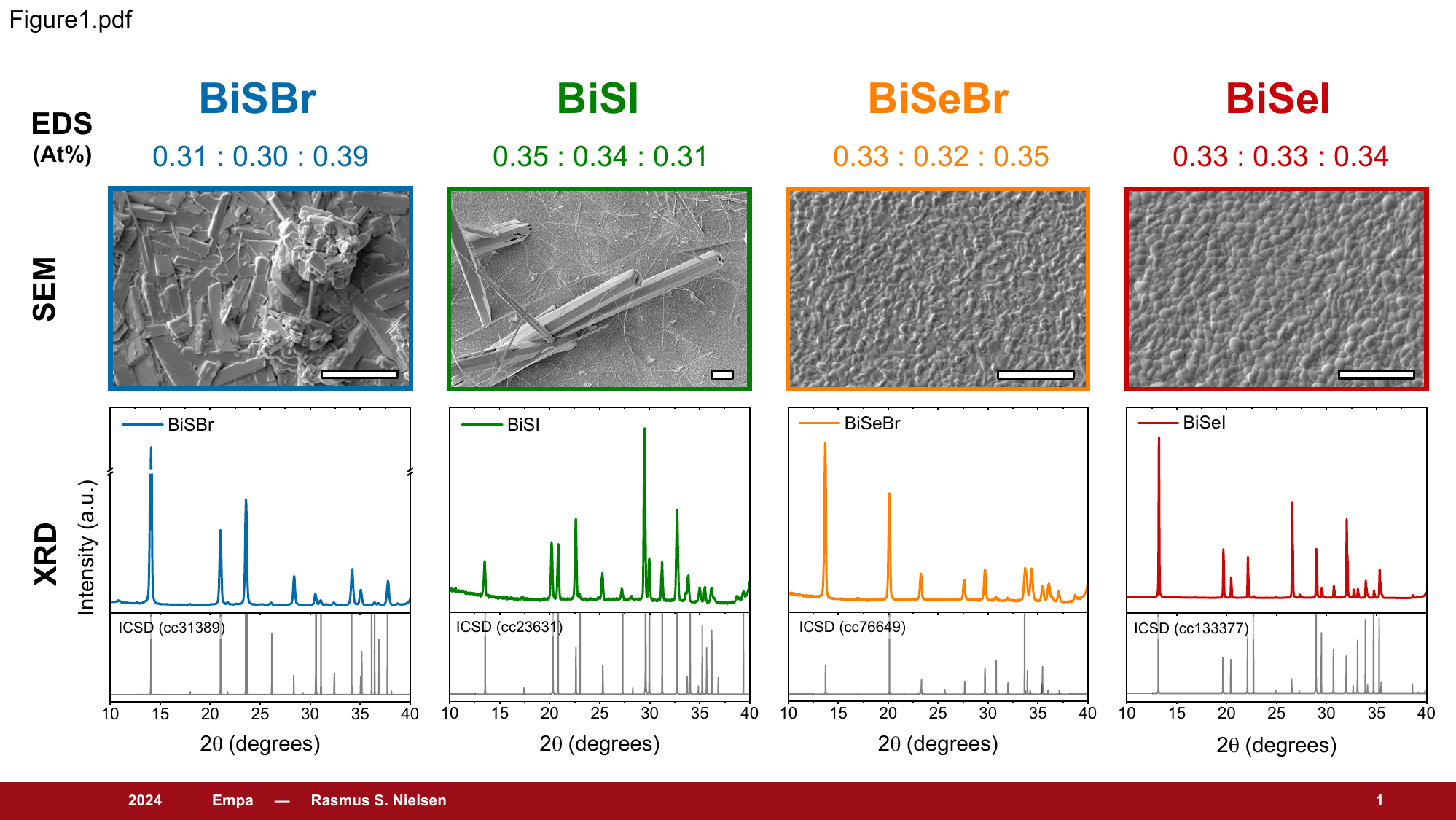}%TRIM=18 40 13 51
    \caption{Structural characterization of Bi-based chalcohalides. The elemental composition was quantified using energy-dispersive X-ray spectroscopy (EDS) mapping and is reported in atomic percent. Top-view scanning electron microscopy (SEM) images depict the microstructural morphology of the samples, with a scale bar of 3 $\mu$m. X-ray diffraction (XRD) patterns are displayed alongside simulated patterns of randomly oriented powders from the best-matching entry in the ICSD database, with the corresponding collection code explicitly stated.}
    \label{fig:Figure1}
\end{figure*}

Scanning electron microscopy (SEM) images were acquired using a Zeiss Gemini 460 equipped with an in-lens secondary electron detector, operated at an acceleration voltage of 2 kV. The SEM was also equipped with an Ultim Max 170 EDS system from Oxford Instruments, which was used to perform energy-dispersive X-ray spectroscopy (EDS) at an acceleration voltage of 10 kV. X-ray diffraction (XRD) patterns were measured using a Bruker D8 Advance in Bragg-Brentano geometry with Cu K$\alpha$ radiation. UV-Vis measurements were conducted using a PerkinElmer Lambda 950 UV/Vis/NIR spectrophotometer equipped with an integrating sphere.

Ab initio calculations based on density functional theory (DFT) \cite{kohn1965a} were carried out as implemented in the VASP code \cite{kresse1996a}. The projector augmented-wave method \cite{kresse1999a} was used to represent the ionic cores \cite{blochl1994a} and for each element the maximum possible number of valence electronic states was considered. Wave functions were represented in a plane-wave basis typically truncated at 600 eV. By using these parameters and a dense k-point grid for Brillouin zone integration of $8\times4\times3$ centered at $\Gamma$, the resulting zero-temperature energies were converged to within 1 meV per formula unit. In the geometry relaxations, a tolerance of 0.005 eV Å$^\text{-1}$ was imposed in the atomic forces.

We employed the Perdew–Burke–Ernzerhof exchange-correlation functional revised for solids \cite{perdew2008a, heyd2003a, schimka2011a}. Long-range dispersion interactions were taken into account through the van der Waals D3 correction scheme \cite{grimme2010a}. Phonon calculations were performed with the small-displacement method and the PHONOPY software \cite{togo2015a}. Large supercells (324 atoms) constructed as $3\times3\times3$ unit cell replications were employed along with a k-point grid of $4\times2\times2$ for sampling of the Brillouin zone.

\normalsize

\section{RESULTS}

First, we characterized the structural properties of the chalcohalide samples synthesized on glass/Mo substrates. We quantified the chemical compositions using EDX mappings, examined microstructural morphologies via top-view SEM imaging, and determined the crystal structures from XRD patterns. The identification of secondary phases is particularly important, as the assessment of intrinsic material properties of a given compound is only meaningful when based on phase-pure samples.

\subsection*{Structural properties of Bi-based chalcohalides}

The structural characterization of the bismuth-based chalcohalide samples is presented in Figure \ref{fig:Figure1}. EDS mapping confirms that all compounds have compositions close to the expected stoichiometric 1:1:1 ratio in the AXY structure, with a slight excess of the halogen species, except for BiSI. SEM images reveal that BiSBr exhibits a platelet morphology with $\sim$3 $\mu$m aggregates, which are more clearly visible in the lower-magnification images in Fig. S1 of the Supplementary Information. In contrast, BiSI forms large needle-like crystals, consistent with previous studies in which BiSI was synthesized using Bridgman-Stockbarger growth \cite{horak1965a, sasaki1965a}, solvothermal \cite{fa2011a}, and hydrothermal methods \cite{su2006a}. For both BiSI and BiSBr, all XRD reflections match the reference patterns from the ICSD database, indicating phase-pure samples. Although these aggregates could correspond to secondary phases, their density and/or long-range order are insufficient for detection via XRD.

\begin{figure*}[t!]
    \centering
    \includegraphics[width=\textwidth,trim={0 0 0 0},clip]{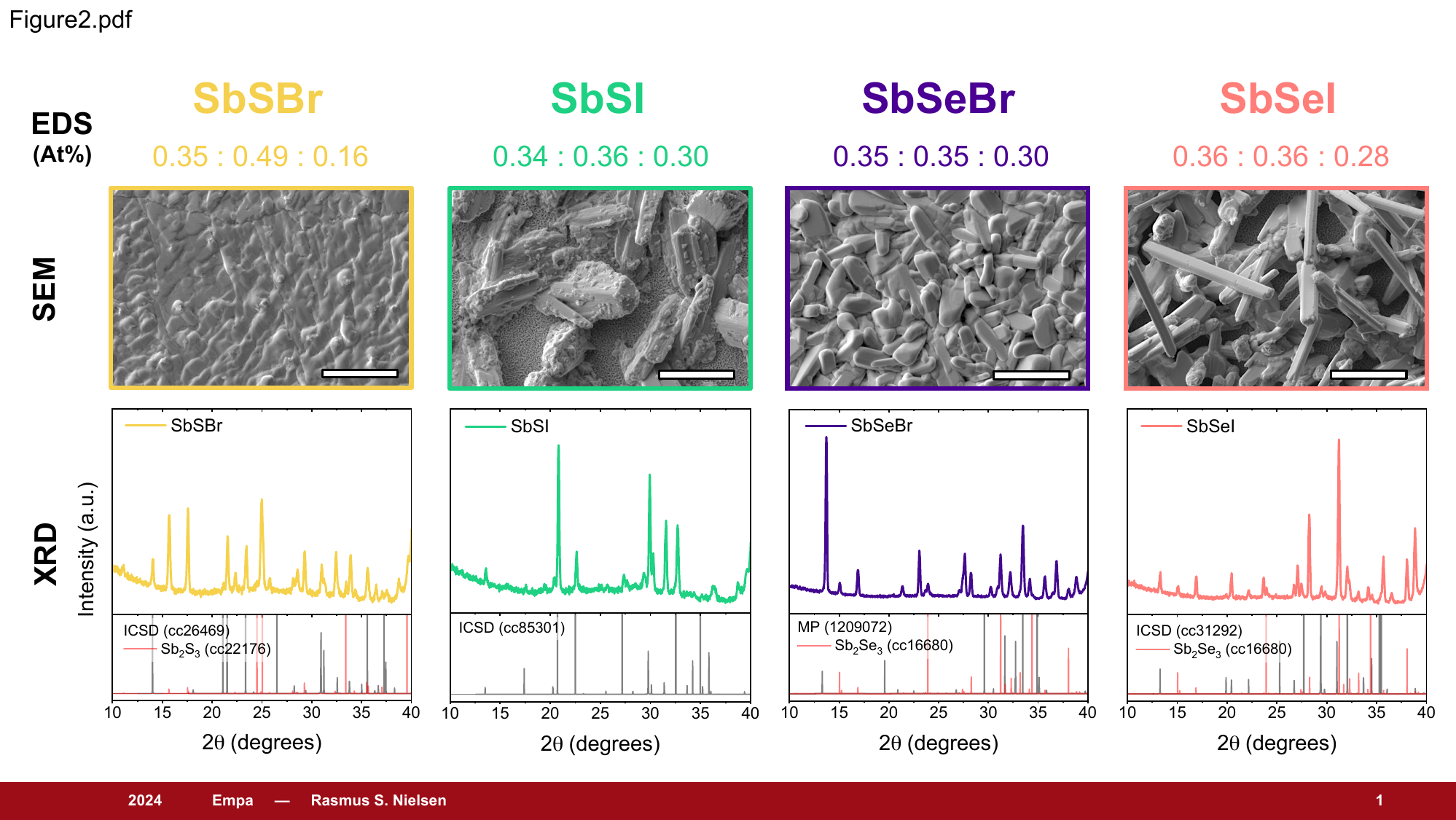}%TRIM=18 40 13 51
    \caption{Structural characterization of Sb-based chalcohalides. The elemental composition was quantified using energy-dispersive X-ray spectroscopy (EDS) mapping and is reported in atomic percent. Top-view scanning electron microscopy (SEM) images depict the microstructural morphology of the samples, with a scale bar of $\mu$m. X-ray diffraction (XRD) patterns are presented alongside simulated patterns for the best-matching ternary phase from the ICSD database, with the corresponding collection code noted. In addition to the desired ternary phase, reference patterns for secondary binary phases are included where corresponding peaks or reflections are observed in the measured XRD patterns.}
    \label{fig:Figure2}
\end{figure*}

The selenides, BiSeBr and BiSeI, exhibit compositions closer to the expected stoichiometry, and their surface morphologies appear more homogeneous compared to their sulfur-based counterparts. As with the sulfides, all XRD reflections match the reference patterns from the ICSD database, suggesting phase-pure samples. Unlike prior reports on BiSeI, which describe crystallographic morphologies as platelet-like structures or rod-like microcrystals obtained via vapor transport and solvothermal synthesis \cite{ganesha1993a, zhu2003a}, our synthesis approach resulted in compact thin films. This distinction is particularly relevant for optoelectronic device applications, where planar architectures are typically used. Given the limited number of studies on BiSeI and BiSeBr, and the fact that we obtained compact thin film morphologies for both of these compounds, investigating their optoelectronic properties is particularly interesting, as these materials can be readily integrated into various optoelectronic applications.

Among the Bi-based compounds, BiSI is the most extensively studied and is the only material to have been explored in various energy-related applications, including solar cells \cite{choi2019a, tiwari2019a, kunioku2016a, hahn2012a}, photocatalysis \cite{zhou2019a}, and photodetectors \cite{farooq2021a}. However, despite its widespread study and use in energy devices, relatively little has been reported on its intrinsic optoelectronic properties. The large needle-like single crystals grown here offer a unique opportunity to study these optoelectronic properties in greater detail.

\subsection*{Structural properties of Sb-based chalcohalides}

The structural characterization of the antimony-based chalcohalide samples is presented in Figure \ref{fig:Figure2}. Unlike the bismuth-based chalcohalides, the antimony-based samples exhibit a compositional deficiency in halogen species, suggesting incomplete halogenation of the binary chalcogenide precursors.  This is further supported by the presence of secondary binary phases in the XRD patterns of all samples, except for SbSI, which contains the highest relative halogen concentration.

SEM images show that only SbSBr forms a compact thin film; however, it is also the most halogen-deficient and exhibits the strongest XRD reflections corresponding to the binary precursor phase. Although reflections corresponding to the desired ternary phase are observed, the lack of phase purity complicates the assessment of intrinsic material properties and their viability for device applications. This underscores the need to further optimize the synthesis process of the Sb-based chalcohalides to achieve complete halogenation of the binary precursors and obtain compact, phase-pure thin films.

\begin{figure*}[t!]
    \centering
    \includegraphics[width=0.8\textwidth,trim={0 0 0 0},clip]{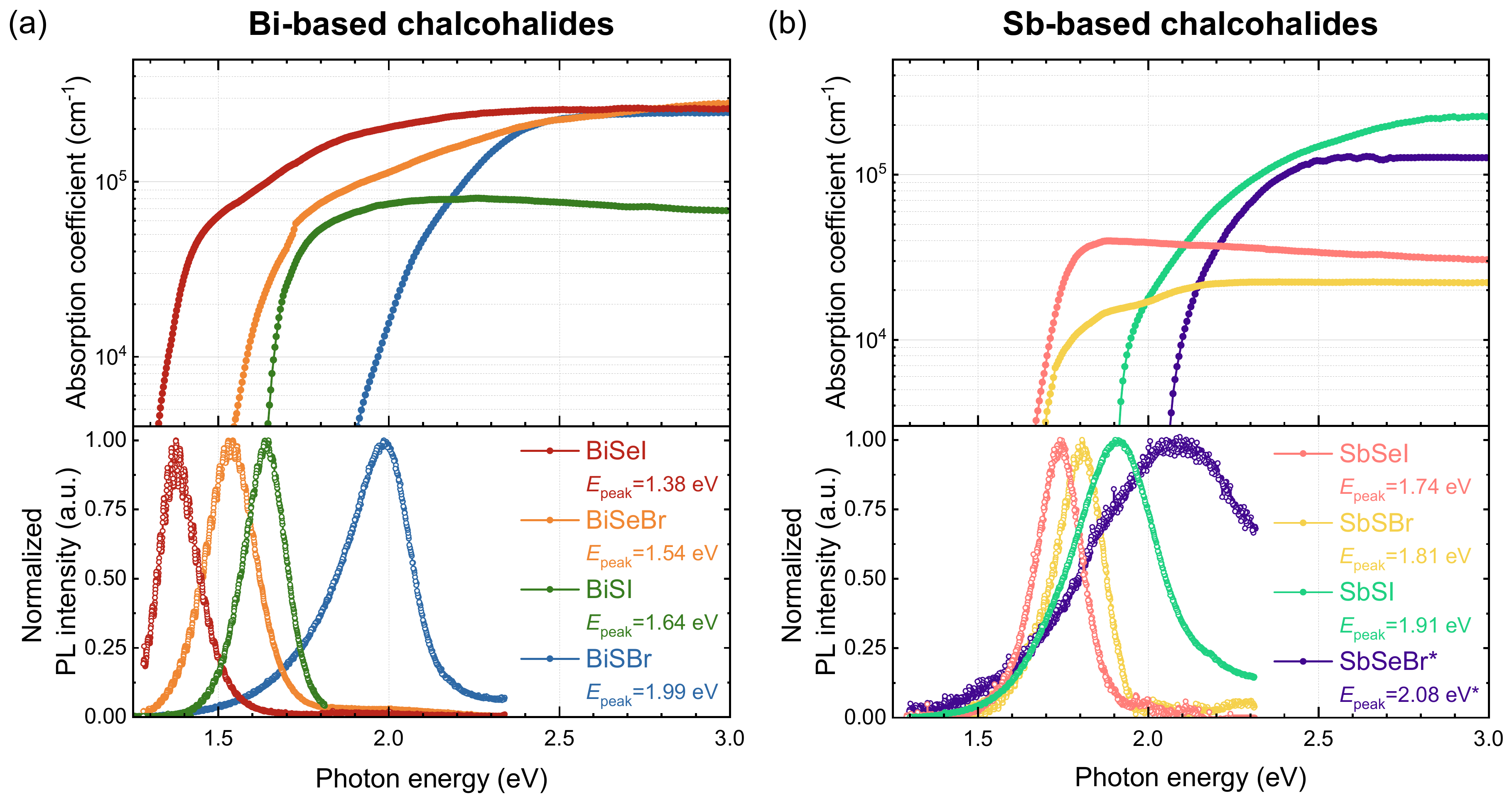}%TRIM=0 0 0 0
    \caption{Absorption coefficient and normalized photoluminescence (PL) intensity for (a) Bi-based and (b) Sb-based chalcohalides. All measurements were carried out at room temperature, with the PL emission peak position of each compound explicitly labeled in the figure. The measurements for SbSeBr are labeled with an asterisk, indicating that the structural determination of this compound was inconclusive, and the observed deviation from the sulfur-selenium trend should be treated with caution.}
    \label{fig:Figure3}
\end{figure*}

It is important to note that no prior reports of SbSeBr exist in the ICSD database. Consequently, the reference powder pattern for this compound was generated using the computationally predicted structure retrieved from the Materials Project \cite{jain2013a}. However, the presence of secondary binary Sb$_\text{2}$Se$_\text{3}$ phases in the XRD patterns and significant uncertainty in the peak identification of the desired ternary phase indicate that results obtained from this sample should be interpreted with caution. Further studies are needed to confirm the structure of SbSeBr and optimize the synthesis process.

Among the Sb-based chalcohalides, SbSI is the most extensively studied. It was the first compound in this family identified as both ferroelectric and photoconductive, with its photoelectric properties investigated by Nitsche and Mertz \cite{nitsche1960a}, and its ferroelectric behavior first reported by Fatuzzo et al. \cite{fatuzzo1962a} The discovery of ferroelectricity in SbSI sparked significant interest, particularly for applications in photodetectors \cite{chen2015a, goedel2016a, sun2019a, shen2020a, ghorpade2023a}, solar cells \cite{choi2018a, nie2018a}, photocatalysts \cite{wang2018a, tasviri2017a, tamilselvan2016a, kwolek2015a}, and nanogenerators \cite{purusothaman2018a}. In contrast, SbSeI, SbSBr, and SbSeBr remain largely unexplored for energy devices, despite their structural similarity to SbSI. Investigating their optoelectronic properties and potential for functional devices could uncover new opportunities for this emerging materials class.

\subsection*{Light absorption and emission}

Following the structural characterization, we proceeded to investigate the light absorption and emission properties of the chalcohalides. To determine the absorption coefficient, we measured both direct and diffuse transmission and reflection using UV-Vis spectroscopy, with a spectrophotometer equipped with an integrating sphere. Since this measurement requires a transparent substrate, we used the chalcohalide samples synthesized on quartz. The absorption coefficient was then calculated using Lambert-Beer’s law. Additionally, we examined photoluminescence (PL) emission at room temperature using a 488 nm excitation laser. Despite observing secondary phases in the Sb-based samples due to incomplete halogenation, we performed UV-Vis and PL spectroscopy on all eight chalcohalide compounds. The results from both techniques are presented in Figure \ref{fig:Figure3}, with panel (a) showing the Bi-based and panel (b) the Sb-based chalcohalides. 

Across all eight chalcohalide compounds, we observe direct bandgaps ranging from 1.38 eV to 2.08 eV, a range that is highly desirable for various semiconductor and optoelectronic applications. While most chalcohalides are known to exhibit indirect fundamental bandgaps \cite{ghorpade2023a}, the energy difference between the indirect and lowest direct transitions is typically so small that the indirect edge has minimal, if any, impact on most optical spectroscopy measurements. The absorption coefficients of the Bi-based compounds reach values on the order of 10$^\text{5}$ cm$^\text{-1}$, highlighting their strong potential as efficient light-absorbing materials.

By comparing the absorption coefficients and PL emission peaks, we find that the absorption onset and PL peak position match closely for both the Bi-based and Sb-based chalcohalides. However, a small anti-Stokes shift is observed in SbSeI and SbSBr, where the emission peak is slightly shifted toward higher photon energies. This shift may be attributed to carrier localization effects, where shallow defect states or compositional inhomogeneities create a preferential recombination pathway at slightly higher energies. Alternatively, excitonic effects could contribute, as variations in exciton binding energies may introduce subtle differences between absorption and emission. The higher bandgap materials, BiSBr and SbSI, exhibit broader and more asymmetric emission peaks, which may arise from a combination of electron-phonon coupling and the influence of defect-related or band tail states. The presence of secondary phases in the Sb-based chalcohalides is not immediately apparent from the absorption and emission characteristics, as both Sb$_\text{2}$Se$_\text{3}$ and Sb$_\text{2}$S$_\text{3}$ have lower bandgaps of approximately 1.2 eV \cite{chen2015b} and 1.7 eV \cite{versavel2007a}, respectively. However, their presence could contribute to the observed peak broadening.

\begin{figure*}[t!]
    \centering
    \includegraphics[width=\textwidth,trim={0 0 0 0},clip]{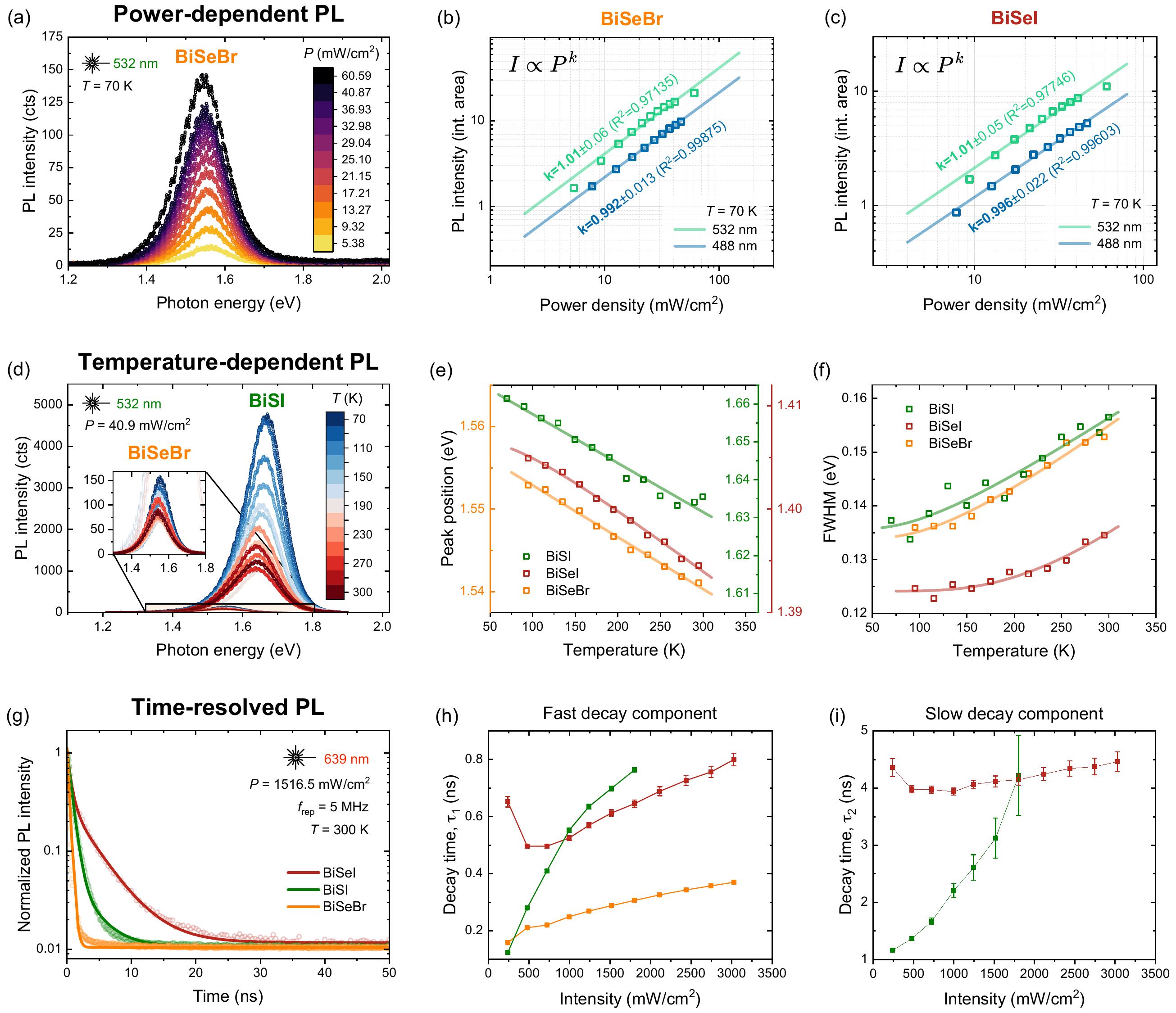}%TRIM=0 0 0 0
    \caption{Photoluminescence characterization of BiSI, BiSeI, and BiSeBr. (a) Power-dependent PL spectrum of BiSeBr at $T=70\,\text{K}$. (b, c) Power dependence of the integrated PL intensity for the two selenide-based thin films, measured using 532 nm and 488 nm excitation lasers, respectively, and fitted to a power-law relation. (d) Temperature-dependent PL spectra of BiSI and BiSeBr (inset), highlighting the significant difference in relative PL yield between the needle-like crystal and the thin films. BiSeI exhibits similar PL yields to BiSeBr (see Figure S3 in the Supplementary Information). (e) Temperature dependence of the fitted PL emission peak positions, modeled using Equation 1. (f) Temperature dependence of the peak width, modeled using Equation 2. (g) Time-resolved PL (TRPL) at room temperature, with transient decays fitted using a bi-exponential function. (h, i) Excitation intensity dependence of the fast and slow decay components, respectively. BiSeBr does not exhibit a slow decay component.}
    \label{fig:Figure4}
\end{figure*}

With the exception of SbSeBr, the substitution of sulfur with selenium generally lowers the bandgap, a trend attributed to variations in the lattice parameters. However, as previously mentioned, the structural determination of SbSeBr was inconclusive, and therefore, the observed deviation from the sulfur-selenium trend should be treated with caution. These measurements may not accurately reflect the intrinsic properties of the desired ternary compound, and as such, the results have been labeled with an asterisk. Since all eight chalcohalide compounds share the same crystal structure, it is also worth exploring solid solutions of both the chalcogenide and halogen anions. Similar to other optoelectronic materials, such as Cu(In,Ga)Se$_\text{2}$ (CIGS) and Cu$_\text{2}$ZnSn(S,Se)$_\text{4}$ (CZTS), where solid solutions have been widely used, chalcohalide solid solutions could provide a means to fine-tune material properties, including the bandgap, further enhancing their versatility for device applications.

\subsection*{Carrier dynamics and electron-phonon interactions}

While room-temperature UV-Vis and PL spectroscopy offer a useful overview of light absorption and emission across the eight chalcohalides, a deeper understanding of their optoelectronic properties requires more comprehensive characterization. To assess carrier recombination dynamics and electron-phonon interactions, we performed power-, temperature-, and time-resolved PL measurements. These measurements were carried out on the three phase-pure chalcohalides selected for their promising morphological and optical characteristics: the BiSeBr and BiSeI thin films, and the needle-like BiSI crystals.

First, we investigated the nature of the emission peaks of the two thin films through power-dependent PL measurements at 70 K using both 488 nm and 532 nm excitation lasers. This analysis helps determine whether the significantly weaker PL intensity observed in the thin films, relative to the needle-like BiSI crystals, arises from excitonic effects or defect-related transitions. Figure \ref{fig:Figure4}(a) shows the power-dependent PL spectra of BiSeBr, while Figures \ref{fig:Figure4}(b) and \ref{fig:Figure4}(c) present the integrated PL intensity, determined by fitting the emission peak with a single Gaussian, as a function of excitation power density for BiSeBr and BiSeI, respectively. The measurements are fitted using the power-law relation $I \propto P^k$, where $I$ is the integrated PL intensity, $P$ is the excitation power density, and $k$ is the power-law exponent.

For both thin films, we consistently obtain $k$ values close to 1 under both excitation wavelengths, indicating that the PL emission peaks originate from band-to-band-like transitions. This behavior rules out defect-mediated recombination, which typically results in sub-linear ($k < 1$) power dependencies, as well as excitonic transitions, which are expected to yield super-linear ($k > 1$) responses \cite{levanyuk1981a, schmidt1992a}.

Having established that the PL emission originates from intrinsic, band-to-band-like transitions, we next investigate the temperature-dependent PL. As shown in Figure \ref{fig:Figure4}(d), the relative PL yield is significantly higher for the needle-like BiSI crystals compared to the BiSeBr and BiSeI thin films (the latter shown in Figure S3 in the Supplementary Information) across all measured temperatures. 

Se-based compounds typically exhibit lower PL yields than their S-based counterparts due to a combination of intrinsic electronic and structural factors. The larger atomic radius and weaker metal–selenium bonds in Se compounds tend to increase the density of structural defects and non-radiative recombination centers, both of which can suppress radiative emission. Recent work by López et al. \cite{lópez2025chalcogenvacanciesrulecharge} demonstrated that Se-based pnictogen chalcohalides (e.g., BiSeI and BiSeBr) exhibit higher equilibrium concentrations of native defects -- particularly selenium vacancies -- than their sulfur-based analogues (e.g., BiSI and BiSBr). This trend is attributed to the lower formation energies and high volatility of selenium under typical synthesis conditions. Furthermore, the reduced bandgaps of Se-based compounds favor thermal population of non-radiative states, which can enhance phonon-assisted decays \cite{zhang2022a, monserrat2018a}. The stronger spin-orbit coupling in selenium, due to its higher atomic mass, can also facilitate exciton relaxation into dark states, further reducing radiative efficiency. These effects have been widely observed in other systems. For instance, monolayer MoSe$_2$ exhibits significantly lower PL intensity than MoS$_2$ under identical conditions, primarily due to increased non-radiative decay pathways and dark exciton formation \cite{robert2020a}. Similarly, CdSe quantum dots without core-shell passivation generally show lower PL quantum yields than CdS counterparts \cite{nguyen2024a}, highlighting a broader trend of diminished PL efficiency in Se-based semiconductors.

This difference in Pl yield may also be partly attributed to a deeper excitation penetration depth in BiSI, coupled with a much higher density of grain boundaries in the thin films, which are expected to act as non-radiative recombination centers. This suggests that the absolute PL intensity and thus the overall optoelectronic quality of the selenide-based thin films could be improved by increasing the crystal grain size or through grain boundary passivation.

To quantify the temperature-dependence of the PL emission, we fit the peak at each temperature with a single Gaussian ($\text{R}^2>0.95$) and track the extracted peak position and spectral width. These trends are shown in Figures \ref{fig:Figure4}(e) and \ref{fig:Figure4}(f), respectively. As both the thermal shift of the bandgap and the broadening of the emission peak are typically attributed to electron-phonon interactions, we model them using Bose-Einstein statistics. The temperature dependence of the bandgap is described by:

\begin{equation}
    E_\mathrm{g}\left(T\right)=E_0- \frac{\lambda}{\exp\left(\hbar\bar{\omega}/k_\mathrm{B}T\right)-1}
\end{equation}

\noindent where $E_0$ is the bandgap at absolute zero, $\lambda$ is a proportionality constant, $\hbar\bar{\omega}$ is the effective phonon energy, and $k_\mathrm{B}T$ is the thermal energy. Similarly, the linewidth broadening is modeled as:

\begin{equation}
    \Gamma_\mathrm{PL}\left(T\right) = \Gamma_0 + \frac{\Gamma_\mathrm{ph}}{\exp \left(\hbar \omega_\mathrm{ph}/k_\mathrm{B}T\right)-1}
\end{equation}

\noindent where $\Gamma_0$ represents the temperature-independent contributions to $\Gamma_\mathrm{PL}$, $\Gamma_\mathrm{ph}$ is the electron-phonon coupling strength, and $\hbar \omega_\mathrm{ph}$ is the characteristic phonon energy. Both models yield reasonably good agreement with the experimental data, and the fitted parameters are summarized in Table \ref{tbl:FittedParameters}.

\begin{table}[t!]
\small
\caption{Extracted parameters from fitting the temperature-dependent PL peak position and linewidth using Bose-Einstein models. $E_\mathrm{g}$ is the room-temperature bandgap, $\lambda$ is the proportionality constant, and $\hbar \bar{\omega}$ is the effective phonon energy associated with the thermal peak shift. $\Gamma_0$ represents the temperature-independent contribution to the linewidth, while $\Gamma_\mathrm{ph}$ and $\hbar \omega_\mathrm{ph}$ describe the electron-phonon coupling strength and the characteristic phonon energy responsible for PL broadening, respectively. BiSeI is marked with an asterisk due to its anomalously large coupling strength and phonon energy, indicating that the simple Bose–Einstein model may not capture the underlying electron-phonon interactions in this sample.}
\vspace{0.25cm}
\label{tbl:FittedParameters}
\begin{tabular*}{1\columnwidth}{@{\extracolsep{\fill}}lccccccc} % Adjusted column alignment
    \hline \vspace{-0.25cm} \\
     & $E_\mathrm{g}$ & $E_0$ & $\lambda$ & $\hbar \bar{\omega}$ & $\Gamma_0$ & $\Gamma_\mathrm{ph}$ & $\hbar \omega_\mathrm{ph}$ \\
    Material & (eV) & (eV) & (meV) & (meV) & (meV) & (meV) & (meV) \vspace{0.05cm}\\
    \hline \vspace{-0.25cm} \\
    BiSeI* & 1.39 & 1.41 & 13.9 & 19.2 & 124 & 161 & 71 \\
    BiSeBr & 1.54 & 1.56 & 0.1 & 0.14 & 134 & 50 & 31 \\
    BiSI \vspace{0.05cm} & 1.63 & 1.67 & 0.1 & 0.06 & 136 & 36 & 26 \\
    \hline 
\end{tabular*}
\end{table}

The temperature-dependent PL behavior of BiSeBr and BiSI is well described by the Bose-Einstein models, as reflected by their modest coupling strengths ($\Gamma_\mathrm{ph} = 50$ meV and $36$ meV, respectively) and characteristic phonon energies ($\hbar \omega_\mathrm{ph} = 31$ meV and $26$ meV). These relatively low values suggest that PL broadening in these materials is primarily governed by moderate electron-phonon interactions, consistent with weakly perturbed radiative recombination. The small values of $\lambda$ and $\hbar \bar{\omega}$ further indicate that the thermal redshift of the PL peak is gradual, and that the underlying phonon modes responsible for broadening are within a physically reasonable energy range. Additionally, the higher bandgap energies (1.54 eV for BiSeBr and 1.63 eV for BiSI) help suppress the thermal activation of non-radiative states, consistent with the defect tolerance of these materials recently reported by López et al. \cite{lópez2025chalcogenvacanciesrulecharge}.

In contrast, the fitted parameters for BiSeI show a significantly higher phonon coupling strength ($\Gamma_\mathrm{ph} = 161$~meV) and an unusually large characteristic phonon energy ($\hbar \omega_\mathrm{ph} = 71$~meV), well beyond the typical energy range of lattice vibrations. This suggests that while the Bose–Einstein model generally describes the temperature dependence of PL in BiSeI, it likely oversimplifies the true broadening mechanisms, potentially involving anharmonic lattice dynamics or multiple phonon modes. These anomalies align with the recent insights into the native defect chemistry of this materials system from López et al. \cite{lópez2025chalcogenvacanciesrulecharge}, who showed that the equilibrium concentration of selenium vacancies (V$_\mathrm{Se}$) in BiSeI is relatively high due to a low formation enthalpy under Se-poor conditions. These vacancies introduce deep charge transition states that drive non-radiative recombination, significantly reducing photocarrier lifetimes and radiative efficiency. Although the capture coefficients of these defects are calculated to be modest, their high concentrations can substantially limit device performance. The detrimental impact of these vacancies may be mitigated through cation-poor synthesis conditions and strategic anion substitutions, offering potential pathways to improve the optoelectronic performance of BiSeI. In contrast, BiSeBr benefits from higher defect formation enthalpies -- particularly for antisite defects -- due to larger ionic radius mismatches, which effectively suppress defect formation. Collectively, these results illustrate that the stronger apparent phonon coupling and enhanced PL broadening observed for BiSeI not only reflect more intense electron–phonon interactions but also highlight the severe impact of high defect densities on non-radiative carrier decay.

To assess whether the carrier dynamics in these compounds reflect similar behavior as suggested by the temperature-dependent PL results, we performed time-resolved photoluminescence (TRPL) measurements, shown in Figure \ref{fig:Figure4}(g). The PL decay in BiSeBr can be modeled using a mono-exponential function, indicating that the carrier dynamics in this sample are relatively straightforward to describe. In contrast, the transient decays for BiSeI and the needle-like BiSI crystal require at least bi-exponential fits, suggesting multiple recombination pathways occurring in parallel and/or carrier localization effects.

To better understand these processes, we performed excitation intensity-dependent TRPL measurements to study how the different decay components evolve with increasing injection levels. The results are shown in Figures \ref{fig:Figure4}(h) and \ref{fig:Figure4}(i) for the fast and slow decay component, respectively. As non-radiative recombination centers become saturated at higher carrier densities, the effective carrier lifetime is expected to increase asymptotically \cite{liu2014a}. This trend is indeed observed for the fast decay component in both BiSI and BiSeBr, suggesting that this component reflects the effective carrier lifetime in these materials. Recent work by Yuan et al. showed that shallow defects typically limit the recombination lifetime in triple-cation perovskites, leading to a decrease in the measured lifetime with increasing carrier concentration \cite{yuan2024a}. Since none of the decay components in any of our samples exhibit this trend, shallow traps are unlikely to be the dominant non-radiative recombination mechanism. This implies that the effective carrier lifetime may be significantly improved by identifying and mitigating deep-level defects.
%https://pubs.acs.org/doi/epdf/10.1021/nl102987z?ref=article_openPDF
%https://www.picoquant.com/applications/category/materials-science/semiconductor-characterization

In contrast, the fast decay component in BiSeI increases approximately linearly with excitation intensity, while the slower component shows little dependence on injection level. This behavior aligns with the temperature-dependent PL results, indicating that the recombination dynamics and electron-phonon interactions in this sample are not well-captured by simple models and point to relatively poor optoelectronic quality. Nevertheless, the fast decay component remains below 1 ns for all three samples, highlighting the need for defect mitigation strategies to enhance carrier lifetimes.

\section{Discussion}

% from (https://onlinelibrary.wiley.com/doi/epdf/10.1002/pssr.202400386?domain=p2p_domain&token=HZ57JZ5FAT8VGQSEKMHU)

\begin{figure}[b!]
    \centering
    \includegraphics[width=\columnwidth,trim={0 0 0 0},clip]{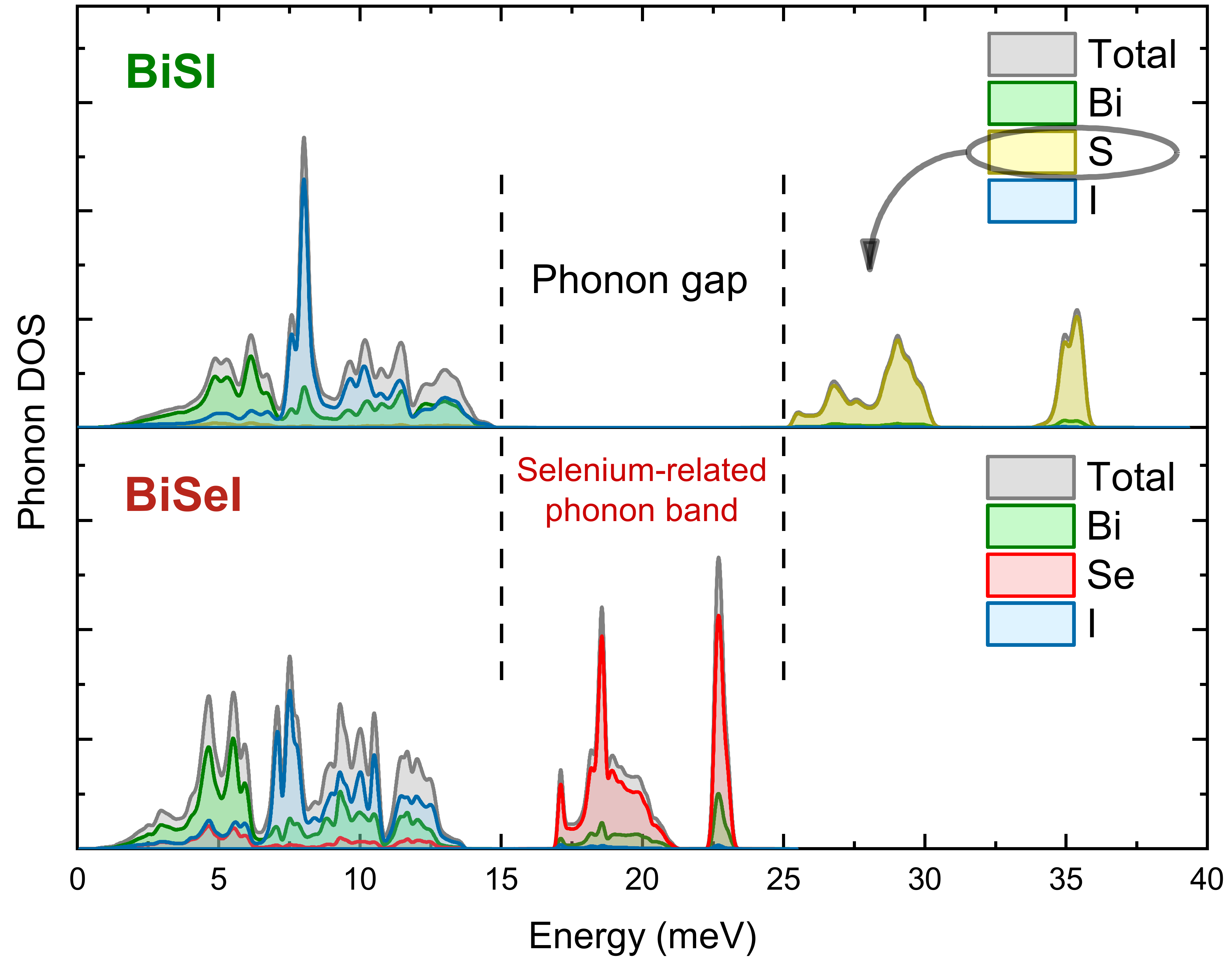}%TRIM=0 0 0 0
    \caption{Total phonon density of states (DOS) as a function of energy for BiSI and BiSeI. In BiSI, the sulfur-related phonon band between 25–30 meV aligns with the characteristic phonon energy responsible for the thermal broadening of the PL emission peak. Immediately below this band is a wide ($\approx$ 10 meV) phonon gap, attributed to the large mass differences between Bi, I, and S atoms. In contrast, BiSeI exhibits a selenium-related phonon band that overlaps the phonon gap region in BiSI, suggesting that a Bi(S,Se)I solid solution could reduce or eliminate this gap. The resulting modification to the phonon structure may suppress electron–phonon interactions and lower the probability of non-radiative recombination.}
    \label{fig:Figure5}
\end{figure}

The optoelectronic behavior of ternary chalcohalides in the (Bi,Sb)(S,Se)(Br,I) family reflects a complex interplay between their electronic structure, defect chemistry, and vibrational dynamics. Among the Bi-based compounds, BiSeBr and BiSI demonstrate favorable characteristics, including moderate electron–phonon interactions and relatively narrow PL linewidths, indicative of efficient radiative recombination. The fitted parameters from the Bose-Einstein models for these compounds -- $\Gamma_\mathrm{ph} = 50$ meV and 36 meV, and $\hbar\omega_\mathrm{ph} = 31$ meV and 26 meV for BiSeBr and BiSI, respectively -- are consistent with typical lattice vibration energies. These values, along with small thermal redshifts of the PL peak, point to phonon-limited broadening that remains manageable for optoelectronic applications. Moreover, their higher bandgaps (1.54–1.63 eV) effectively suppress thermal activation of non-radiative states, supporting their strong PL emission.

In contrast, BiSeI stands out for its anomalously strong phonon coupling ($\Gamma_\mathrm{ph} = 161$ meV) and a fitted characteristic phonon energy ($\hbar\omega_\mathrm{ph} = 71$ meV) that is well beyond the physical range of known vibrational modes. While the Bose-Einstein model captures the general temperature trend, these values suggest that the model oversimplifies the underlying recombination mechanisms. The exaggerated coupling suggests a combination of anharmonic lattice effects, multi-mode phonon interactions, and substantial carrier trapping by deep defects.

Recent defect calculations by López et al.\cite{lópez2025chalcogenvacanciesrulecharge} shed light on the native defect chemistry, identifying chalcogen vacancies as the killer point defects in these systems. BiSeI is particularly susceptible to form Se vacancies (V$_\mathrm{Se}$) due to a relatively low formation enthalpy under Se-poor conditions. These vacancies act as deep non-radiative centers, consistent with the poor PL performance and sub-nanosecond carrier lifetimes observed in our measurements. In contrast, BiSeBr exhibits higher antisite formation energies, partially attributed to larger ionic radius mismatches, which enhance its defect tolerance and radiative efficiency.

TRPL measurements further confirm these findings. BiSeBr exhibits clean mono-exponential decay, indicative of a single dominant recombination pathway consistent with efficient band-to-band emission. In contrast, BiSeI and BiSI show bi-exponential decay, suggesting the presence of multiple recombination channels and possible carrier localization. In BiSeI, the fast decay component increases linearly with excitation density and remains below 1 ns across all measured conditions, consistent with a high density of unsaturable non-radiative traps. BiSeBr and BiSI, however, exhibit an asymptotic increase in their fast decay components, consistent with saturation of defects. These trends support the notion that deep-level defects govern recombination in BiSeI, while BiSeBr may be further optimized through surface or grain boundary passivation.

In addition to defect chemistry, vibrational effects also play a central role in the optoelectronic performance of these chalcohalides. Figure \ref{fig:Figure5} shows the total phonon density of states (DOS) for BiSI, calculated using density functional theory (DFT), revealing a pronounced phonon gap just below the sulfur-related phonon band (25–30 meV), which closely aligns with the experimentally extracted characteristic phonon energy $\hbar\omega_\mathrm{ph}$. This gap originates from the large atomic mass differences between the heavy pnictogen and the much lighter sulfur atoms and is consistently observed across all sulfide-based compounds (Figure S4 in the Supplementary Information). While this gap can reduce phonon scattering pathways -- potentially improving radiative efficiency -- it may also enhance energy dissipation through specific high-energy modes, promoting non-radiative recombination. In contrast, the corresponding selenium-based compounds lack this phonon gap, as selenium-related phonon modes extend into the previously forbidden vibrational range. This filled gap intensifies carrier–phonon scattering, providing a compelling vibrational explanation for the anomalous electron-phonon coupling strength observed in BiSeI.

Forming solid solutions, such as Bi(S,Se)I, may help mitigate these issues. Partial substitution of S with Se or Br with I offers a means to engineer the phonon landscape, reintroducing beneficial gaps or shifting scattering bands to suppress non-radiative recombination pathways. Notably, López et al.\cite{lópez2025chalcogenvacanciesrulecharge} reported that such substitutions also improve intrinsic point defect tolerance, providing a dual strategy for optimizing optoelectronic performance through combined phonon and defect engineering.

Altogether, these results highlight the importance of both intrinsic material design and extrinsic processing conditions in tuning the optoelectronic properties of chalcohalides. While BiSeI is currently limited by strong electron-phonon coupling and a high equilibrium concentration of native defects, BiSeBr and BiSI present promising platforms for further development. Strategies such as targeted alloying, grain boundary passivation, and optimization of growth conditions will be key to advancing this materials class toward high-efficiency applications in photovoltaics and photodetection.

\section{Conclusion}

In summary, we have comprehensively investigated the optoelectronic properties of Bi- and Sb-based chalcohalides, revealing key structure-function relationships within this emerging material class. Structural characterization confirmed that our novel two-step PVD synthesis reliably produces phase-pure Bi-based compounds, while the Sb-based counterparts require further optimization to achieve fully halogenated samples. Through power-, temperature-dependent and time-resolved PL (TRPL) measurements, we demonstrated that BiSeBr and BiSI exhibit favorable radiative characteristics, including band-to-band like transitions, moderate electron–phonon coupling and relatively narrow PL linewidths. These properties were in part attributed to their higher bandgaps, which suppress thermal activation of non-radiative states, and higher defect formation energies, which limit the equilibrium concentration of deep-level recombination centers. In contrast, BiSeI showed strong phonon coupling, anomalous PL broadening, and rapid non-radiative decay due to a complex defect landscape localizing and killing photoexcited charge carriers. These findings were further supported by first-principles defect calculations and phonon dispersion analysis, which revealed a phonon gap in all sulfide-based compounds that reduces scattering pathways, but is filled by selenium-related modes in the selenide-based compounds, which is predicted to suppress electron-phonon interactions and enhance carrier lifetimes.

Our results establish a clear structure–property relationship within this material family, where defect formation energies and phonon dynamics play a critical role in determining optoelectronic performance. We suggest that BiSeBr and BiSI hold promise for further development, while BiSeI requires defect management to improve its radiative efficiency. Strategies such as solid-solutions engineering (e.g., Bi(S,Se)I), optimizing growth conditions, and implementing surface or grain boundary passivation can mitigate defect-induced losses and tune the native phonon landscape. By demonstrating how intrinsic defect chemistry and lattice dynamics govern light emission and carrier lifetimes, this study provides a roadmap for optimizing chalcohalides for high-performance applications in photovoltaics, photodetectors, and beyond.\\

\section*{Conflicts of interest}
There are no conflicts of interest to declare.\\

\section*{Acknowledgements}
The work presented here is supported by the Carlsberg Foundation, grant CF24-0200. This project received funding from the European Union’s Horizon research and innovation programme under grant agreements number 866018 (SENSATE); by the Spanish Ministry of Science and Innovation project number PCI2023-145971-2 (ACT-FAST) from the CETP-Partnership Program 2022; and by the Cost Association project number CA-706 21148 (Renew-PV). This work is also part of Maria de Maeztu Units of Excellence Programme CEX2023-001300-M, funded by MICIU/AEI/10.13039/501100011033. The authors from UPC and IREC belong to the Micro and Nanotechnologies for Solar Energy 709 Group (MNTSolar) Consolidated Research Group of the “Generalitat de Catalunya” (2021 SGR 01286). E.S. acknowledges the ICREA Academia program. I.C., acknowledges Inno-PV (PID2022-140226OB-C31, PID2022-140226OB-C32), funded by MCIN/AEI/10.13039/501100 011033/FEDER. C.L. acknowledges support from the Spanish Ministry of Science, Innovation and Universities under an FPU grant. C.C. acknowledges support by MICIN/AEI/10.13039/501100011033 and ERDF/EU under the grants PID2023-146623NB-I00 and PID2023-147469NB-C21 and by the Generalitat de Catalunya under the grants 2021SGR-00343, 2021SGR-01519 and 2021SGR-01411. Computational support was provided by the Red Española de Supercomputación under the grants FI-2024-1-0005, FI-2024-2-0003, FI-2024-3-0004, FI-2024-1-0025, FI-2024-2-0006, FI-2025-1-0015 and FI-2025-1-0001.\\

\section*{Data availability}
All relevant data and code for analysis generated in the course of this work is freely available at \url{https://doi.org/10.5281/zenodo.15614394}.\\


\nocite{*}

\bibliography{references}% Produces the bibliography via BibTeX.

\newcommand{\singleletter}[1]{#1}
\begin{thebibliography}{78}%
\makeatletter
\providecommand \@ifxundefined [1]{%
 \@ifx{#1\undefined}
}%
\providecommand \@ifnum [1]{%
 \ifnum #1\expandafter \@firstoftwo
 \else \expandafter \@secondoftwo
 \fi
}%
\providecommand \@ifx [1]{%
 \ifx #1\expandafter \@firstoftwo
 \else \expandafter \@secondoftwo
 \fi
}%
\providecommand \natexlab [1]{#1}%
\providecommand \enquote  [1]{``#1''}%
\providecommand \bibnamefont  [1]{#1}%
\providecommand \bibfnamefont [1]{#1}%
\providecommand \citenamefont [1]{#1}%
\providecommand \href@noop [0]{\@secondoftwo}%
\providecommand \href [0]{\begingroup \@sanitize@url \@href}%
\providecommand \@href[1]{\@@startlink{#1}\@@href}%
\providecommand \@@href[1]{\endgroup#1\@@endlink}%
\providecommand \@sanitize@url [0]{\catcode `\\12\catcode `\$12\catcode `\&12\catcode `\#12\catcode `\^12\catcode `\_12\catcode `\%12\relax}%
\providecommand \@@startlink[1]{}%
\providecommand \@@endlink[0]{}%
\providecommand \url  [0]{\begingroup\@sanitize@url \@url }%
\providecommand \@url [1]{\endgroup\@href {#1}{\urlprefix }}%
\providecommand \urlprefix  [0]{URL }%
\providecommand \Eprint [0]{\href }%
\providecommand \doibase [0]{https://doi.org/}%
\providecommand \selectlanguage [0]{\@gobble}%
\providecommand \bibinfo  [0]{\@secondoftwo}%
\providecommand \bibfield  [0]{\@secondoftwo}%
\providecommand \translation [1]{[#1]}%
\providecommand \BibitemOpen [0]{}%
\providecommand \bibitemStop [0]{}%
\providecommand \bibitemNoStop [0]{.\EOS\space}%
\providecommand \EOS [0]{\spacefactor3000\relax}%
\providecommand \BibitemShut  [1]{\csname bibitem#1\endcsname}%
\let\auto@bib@innerbib\@empty
%</preamble>
\bibitem [{\citenamefont {Wang}\ \emph {et~al.}(2021)\citenamefont {Wang}, \citenamefont {Huang}, \citenamefont {Xue}, \citenamefont {Tong}, \citenamefont {Zhu},\ and\ \citenamefont {Yang}}]{wang2021a}%
  \BibitemOpen
  \bibfield  {author} {\bibinfo {author} {\bibfnamefont {R.}~\bibnamefont {Wang}}, \bibinfo {author} {\bibfnamefont {T.}~\bibnamefont {Huang}}, \bibinfo {author} {\bibfnamefont {J.}~\bibnamefont {Xue}}, \bibinfo {author} {\bibfnamefont {J.}~\bibnamefont {Tong}}, \bibinfo {author} {\bibfnamefont {K.}~\bibnamefont {Zhu}},\ and\ \bibinfo {author} {\bibfnamefont {Y.}~\bibnamefont {Yang}},\ }\bibfield  {title} {\bibinfo {title} {\singleletter{Prospects for metal halide perovskite-based tandem solar cells}},\ }\href {https://doi.org/10.1038/s41566-021-00809-8} {\bibfield  {journal} {\bibinfo  {journal} {Nature Photonics}\ }\textbf {\bibinfo {volume} {15}},\ \bibinfo {pages} {411} (\bibinfo {year} {2021})}\BibitemShut {NoStop}%
\bibitem [{\citenamefont {Liu}\ \emph {et~al.}(2021{\natexlab{a}})\citenamefont {Liu}, \citenamefont {Xu}, \citenamefont {Bai}, \citenamefont {Jin}, \citenamefont {Wang}, \citenamefont {Friend},\ and\ \citenamefont {Gao}}]{liu2021a}%
  \BibitemOpen
  \bibfield  {author} {\bibinfo {author} {\bibfnamefont {X.~K.}\ \bibnamefont {Liu}}, \bibinfo {author} {\bibfnamefont {W.}~\bibnamefont {Xu}}, \bibinfo {author} {\bibfnamefont {S.}~\bibnamefont {Bai}}, \bibinfo {author} {\bibfnamefont {Y.}~\bibnamefont {Jin}}, \bibinfo {author} {\bibfnamefont {J.}~\bibnamefont {Wang}}, \bibinfo {author} {\bibfnamefont {R.~H.}\ \bibnamefont {Friend}},\ and\ \bibinfo {author} {\bibfnamefont {F.}~\bibnamefont {Gao}},\ }\bibfield  {title} {\bibinfo {title} {\singleletter{Metal halide perovskites for light-emitting diodes}},\ }\href {https://doi.org/10.1038/s41563-020-0784-7} {\bibfield  {journal} {\bibinfo  {journal} {Nature Materials}\ }\textbf {\bibinfo {volume} {20}},\ \bibinfo {pages} {10} (\bibinfo {year} {2021}{\natexlab{a}})}\BibitemShut {NoStop}%
\bibitem [{\citenamefont {Ren}\ \emph {et~al.}(2022)\citenamefont {Ren}, \citenamefont {Yue}, \citenamefont {Li}, \citenamefont {Fang}, \citenamefont {Gasem}, \citenamefont {Leszczynski}, \citenamefont {Qu}, \citenamefont {Wang},\ and\ \citenamefont {Fan}}]{ren2022a}%
  \BibitemOpen
  \bibfield  {author} {\bibinfo {author} {\bibfnamefont {K.}~\bibnamefont {Ren}}, \bibinfo {author} {\bibfnamefont {S.}~\bibnamefont {Yue}}, \bibinfo {author} {\bibfnamefont {C.}~\bibnamefont {Li}}, \bibinfo {author} {\bibfnamefont {Z.}~\bibnamefont {Fang}}, \bibinfo {author} {\bibfnamefont {K.~A.}\ \bibnamefont {Gasem}}, \bibinfo {author} {\bibfnamefont {J.}~\bibnamefont {Leszczynski}}, \bibinfo {author} {\bibfnamefont {S.}~\bibnamefont {Qu}}, \bibinfo {author} {\bibfnamefont {Z.}~\bibnamefont {Wang}},\ and\ \bibinfo {author} {\bibfnamefont {M.}~\bibnamefont {Fan}},\ }\bibfield  {title} {\bibinfo {title} {\singleletter{Metal halide perovskites for photocatalysis applications}},\ }\href {https://doi.org/10.1039/d1ta09148d} {\bibfield  {journal} {\bibinfo  {journal} {Journal of Materials Chemistry A}\ }\textbf {\bibinfo {volume} {10}},\ \bibinfo {pages} {407} (\bibinfo {year} {2022})}\BibitemShut {NoStop}%
\bibitem [{\citenamefont {Lei}\ \emph {et~al.}(2021)\citenamefont {Lei}, \citenamefont {Dong}, \citenamefont {Gundogdu},\ and\ \citenamefont {So}}]{lei2021a}%
  \BibitemOpen
  \bibfield  {author} {\bibinfo {author} {\bibfnamefont {L.}~\bibnamefont {Lei}}, \bibinfo {author} {\bibfnamefont {Q.}~\bibnamefont {Dong}}, \bibinfo {author} {\bibfnamefont {K.}~\bibnamefont {Gundogdu}},\ and\ \bibinfo {author} {\bibfnamefont {F.}~\bibnamefont {So}},\ }\bibfield  {title} {\bibinfo {title} {\singleletter{Metal Halide Perovskites for Laser Applications}},\ }\href {https://doi.org/10.1002/adfm.202010144} {\bibfield  {journal} {\bibinfo  {journal} {Advanced Functional Materials}\ }\textbf {\bibinfo {volume} {31}},\ \bibinfo {pages} {2010144} (\bibinfo {year} {2021})}\BibitemShut {NoStop}%
\bibitem [{\citenamefont {Cheng}\ \emph {et~al.}(2022)\citenamefont {Cheng}, \citenamefont {Peng}, \citenamefont {Jen},\ and\ \citenamefont {Yip}}]{cheng2022a}%
  \BibitemOpen
  \bibfield  {author} {\bibinfo {author} {\bibfnamefont {Y.}~\bibnamefont {Cheng}}, \bibinfo {author} {\bibfnamefont {Y.}~\bibnamefont {Peng}}, \bibinfo {author} {\bibfnamefont {A.~K.}\ \bibnamefont {Jen}},\ and\ \bibinfo {author} {\bibfnamefont {H.~L.}\ \bibnamefont {Yip}},\ }\bibfield  {title} {\bibinfo {title} {\singleletter{Development and Challenges of Metal Halide Perovskite Solar Modules}},\ }\href {https://doi.org/10.1002/solr.202100545} {\bibfield  {journal} {\bibinfo  {journal} {Solar RRL}\ }\textbf {\bibinfo {volume} {6}},\ \bibinfo {pages} {2100545} (\bibinfo {year} {2022})}\BibitemShut {NoStop}%
\bibitem [{\citenamefont {De~Wolf}\ \emph {et~al.}(2014)\citenamefont {De~Wolf}, \citenamefont {Holovsky}, \citenamefont {Moon}, \citenamefont {Löper}, \citenamefont {Niesen}, \citenamefont {Ledinsky}, \citenamefont {Haug}, \citenamefont {Yum},\ and\ \citenamefont {Ballif}}]{de2014a}%
  \BibitemOpen
  \bibfield  {author} {\bibinfo {author} {\bibfnamefont {S.}~\bibnamefont {De~Wolf}}, \bibinfo {author} {\bibfnamefont {J.}~\bibnamefont {Holovsky}}, \bibinfo {author} {\bibfnamefont {S.-J.}\ \bibnamefont {Moon}}, \bibinfo {author} {\bibfnamefont {P.}~\bibnamefont {Löper}}, \bibinfo {author} {\bibfnamefont {B.}~\bibnamefont {Niesen}}, \bibinfo {author} {\bibfnamefont {M.}~\bibnamefont {Ledinsky}}, \bibinfo {author} {\bibfnamefont {F.-J.}\ \bibnamefont {Haug}}, \bibinfo {author} {\bibfnamefont {J.-H.}\ \bibnamefont {Yum}},\ and\ \bibinfo {author} {\bibfnamefont {C.}~\bibnamefont {Ballif}},\ }\bibfield  {title} {\bibinfo {title} {\singleletter{Organometallic Halide Perovskites: Sharp Optical Absorption Edge and Its Relation to Photovoltaic Performance}},\ }\href {https://doi.org/10.1021/jz500279b} {\bibfield  {journal} {\bibinfo  {journal} {Journal of Physical Chemistry Letters}\ }\textbf {\bibinfo {volume} {5}},\ \bibinfo {pages} {1035} (\bibinfo {year} {2014})}\BibitemShut {NoStop}%
\bibitem [{\citenamefont {Stranks}\ \emph {et~al.}(2013)\citenamefont {Stranks}, \citenamefont {Eperon}, \citenamefont {Grancini}, \citenamefont {Menelaou}, \citenamefont {Alcocer}, \citenamefont {Leijtens}, \citenamefont {Herz}, \citenamefont {Petrozza},\ and\ \citenamefont {Snaith}}]{stranks2013a}%
  \BibitemOpen
  \bibfield  {author} {\bibinfo {author} {\bibfnamefont {S.~D.}\ \bibnamefont {Stranks}}, \bibinfo {author} {\bibfnamefont {G.~E.}\ \bibnamefont {Eperon}}, \bibinfo {author} {\bibfnamefont {G.}~\bibnamefont {Grancini}}, \bibinfo {author} {\bibfnamefont {C.}~\bibnamefont {Menelaou}}, \bibinfo {author} {\bibfnamefont {M.~J.}\ \bibnamefont {Alcocer}}, \bibinfo {author} {\bibfnamefont {T.}~\bibnamefont {Leijtens}}, \bibinfo {author} {\bibfnamefont {L.~M.}\ \bibnamefont {Herz}}, \bibinfo {author} {\bibfnamefont {A.}~\bibnamefont {Petrozza}},\ and\ \bibinfo {author} {\bibfnamefont {H.~J.}\ \bibnamefont {Snaith}},\ }\bibfield  {title} {\bibinfo {title} {\singleletter{Electron-hole diffusion lengths exceeding 1 micrometer in an organometal trihalide perovskite absorber}},\ }\href {https://doi.org/10.1126/science.1243982} {\bibfield  {journal} {\bibinfo  {journal} {Science}\ }\textbf {\bibinfo {volume} {342}},\ \bibinfo {pages} {341} (\bibinfo {year} {2013})}\BibitemShut {NoStop}%
\bibitem [{\citenamefont {deQuilettes}\ \emph {et~al.}(2016)\citenamefont {deQuilettes}, \citenamefont {Koch}, \citenamefont {Burke}, \citenamefont {Paranji}, \citenamefont {Shropshire}, \citenamefont {Ziffer},\ and\ \citenamefont {Ginger}}]{dequilettes2016a}%
  \BibitemOpen
  \bibfield  {author} {\bibinfo {author} {\bibfnamefont {D.~W.}\ \bibnamefont {deQuilettes}}, \bibinfo {author} {\bibfnamefont {S.}~\bibnamefont {Koch}}, \bibinfo {author} {\bibfnamefont {S.}~\bibnamefont {Burke}}, \bibinfo {author} {\bibfnamefont {R.~K.}\ \bibnamefont {Paranji}}, \bibinfo {author} {\bibfnamefont {A.~J.}\ \bibnamefont {Shropshire}}, \bibinfo {author} {\bibfnamefont {M.~E.}\ \bibnamefont {Ziffer}},\ and\ \bibinfo {author} {\bibfnamefont {D.~S.}\ \bibnamefont {Ginger}},\ }\bibfield  {title} {\bibinfo {title} {\singleletter{Photoluminescence Lifetimes Exceeding 8 $\mu$s and Quantum Yields Exceeding 30\% in Hybrid Perovskite Thin Films by Ligand Passivation}},\ }\href {https://doi.org/10.1021/acsenergylett.6b00236} {\bibfield  {journal} {\bibinfo  {journal} {ACS Energy Letters}\ }\textbf {\bibinfo {volume} {1}},\ \bibinfo {pages} {438} (\bibinfo {year} {2016})}\BibitemShut {NoStop}%
\bibitem [{\citenamefont {Babayigit}\ \emph {et~al.}(2016)\citenamefont {Babayigit}, \citenamefont {Duy~Thanh}, \citenamefont {Ethirajan}, \citenamefont {Manca}, \citenamefont {Muller}, \citenamefont {Boyen},\ and\ \citenamefont {Conings}}]{babayigit2016a}%
  \BibitemOpen
  \bibfield  {author} {\bibinfo {author} {\bibfnamefont {A.}~\bibnamefont {Babayigit}}, \bibinfo {author} {\bibfnamefont {D.}~\bibnamefont {Duy~Thanh}}, \bibinfo {author} {\bibfnamefont {A.}~\bibnamefont {Ethirajan}}, \bibinfo {author} {\bibfnamefont {J.}~\bibnamefont {Manca}}, \bibinfo {author} {\bibfnamefont {M.}~\bibnamefont {Muller}}, \bibinfo {author} {\bibfnamefont {H.~G.}\ \bibnamefont {Boyen}},\ and\ \bibinfo {author} {\bibfnamefont {B.}~\bibnamefont {Conings}},\ }\bibfield  {title} {\bibinfo {title} {\singleletter{Assessing the toxicity of Pb-and Sn-based perovskite solar cells in model organism Danio rerio}},\ }\href {https://doi.org/10.1038/srep18721} {\bibfield  {journal} {\bibinfo  {journal} {Scientific Reports}\ }\textbf {\bibinfo {volume} {6}},\ \bibinfo {pages} {18721} (\bibinfo {year} {2016})}\BibitemShut {NoStop}%
\bibitem [{\citenamefont {Serrano-Luján}\ \emph {et~al.}(2015)\citenamefont {Serrano-Luján}, \citenamefont {Espinosa~Martinez}, \citenamefont {Larsen-Olsen}, \citenamefont {Abad}, \citenamefont {Urbina},\ and\ \citenamefont {Krebs}}]{serrano2015a}%
  \BibitemOpen
  \bibfield  {author} {\bibinfo {author} {\bibfnamefont {L.}~\bibnamefont {Serrano-Luján}}, \bibinfo {author} {\bibfnamefont {N.}~\bibnamefont {Espinosa~Martinez}}, \bibinfo {author} {\bibfnamefont {T.~T.}\ \bibnamefont {Larsen-Olsen}}, \bibinfo {author} {\bibfnamefont {J.}~\bibnamefont {Abad}}, \bibinfo {author} {\bibfnamefont {A.}~\bibnamefont {Urbina}},\ and\ \bibinfo {author} {\bibfnamefont {F.~C.}\ \bibnamefont {Krebs}},\ }\bibfield  {title} {\bibinfo {title} {\singleletter{Tin- and Lead-Based Perovskite Solar Cells under Scrutiny: An Environmental Perspective}},\ }\href {https://doi.org/10.1002/aenm.201501119} {\bibfield  {journal} {\bibinfo  {journal} {Advanced Energy Materials}\ }\textbf {\bibinfo {volume} {5}},\ \bibinfo {pages} {1501119} (\bibinfo {year} {2015})}\BibitemShut {NoStop}%
\bibitem [{\citenamefont {Duan}\ \emph {et~al.}(2023)\citenamefont {Duan}, \citenamefont {Walter}, \citenamefont {Chang}, \citenamefont {Bullock}, \citenamefont {Kang}, \citenamefont {Phang}, \citenamefont {Weber}, \citenamefont {White}, \citenamefont {Macdonald}, \citenamefont {Catchpole},\ and\ \citenamefont {Shen}}]{duan2023a}%
  \BibitemOpen
  \bibfield  {author} {\bibinfo {author} {\bibfnamefont {L.}~\bibnamefont {Duan}}, \bibinfo {author} {\bibfnamefont {D.}~\bibnamefont {Walter}}, \bibinfo {author} {\bibfnamefont {N.}~\bibnamefont {Chang}}, \bibinfo {author} {\bibfnamefont {J.}~\bibnamefont {Bullock}}, \bibinfo {author} {\bibfnamefont {D.}~\bibnamefont {Kang}}, \bibinfo {author} {\bibfnamefont {S.~P.}\ \bibnamefont {Phang}}, \bibinfo {author} {\bibfnamefont {K.}~\bibnamefont {Weber}}, \bibinfo {author} {\bibfnamefont {T.}~\bibnamefont {White}}, \bibinfo {author} {\bibfnamefont {D.}~\bibnamefont {Macdonald}}, \bibinfo {author} {\bibfnamefont {K.}~\bibnamefont {Catchpole}},\ and\ \bibinfo {author} {\bibfnamefont {H.}~\bibnamefont {Shen}},\ }\bibfield  {title} {\bibinfo {title} {\singleletter{Stability challenges for the commercialization of perovskite–silicon tandem solar cells}},\ }\href {https://doi.org/10.1038/s41578-022-00521-1} {\bibfield  {journal} {\bibinfo  {journal} {Nature Reviews Materials}\ }\textbf {\bibinfo {volume} {8}},\
  \bibinfo {pages} {261} (\bibinfo {year} {2023})}\BibitemShut {NoStop}%
\bibitem [{\citenamefont {Conings}\ \emph {et~al.}(2015)\citenamefont {Conings}, \citenamefont {Drijkoningen}, \citenamefont {Gauquelin}, \citenamefont {Babayigit}, \citenamefont {D'Haen}, \citenamefont {D'Olieslaeger}, \citenamefont {Ethirajan}, \citenamefont {Verbeeck}, \citenamefont {Manca}, \citenamefont {Mosconi}, \citenamefont {De~Angelis},\ and\ \citenamefont {Boyen}}]{conings2015a}%
  \BibitemOpen
  \bibfield  {author} {\bibinfo {author} {\bibfnamefont {B.}~\bibnamefont {Conings}}, \bibinfo {author} {\bibfnamefont {J.}~\bibnamefont {Drijkoningen}}, \bibinfo {author} {\bibfnamefont {N.}~\bibnamefont {Gauquelin}}, \bibinfo {author} {\bibfnamefont {A.}~\bibnamefont {Babayigit}}, \bibinfo {author} {\bibfnamefont {J.}~\bibnamefont {D'Haen}}, \bibinfo {author} {\bibfnamefont {L.}~\bibnamefont {D'Olieslaeger}}, \bibinfo {author} {\bibfnamefont {A.}~\bibnamefont {Ethirajan}}, \bibinfo {author} {\bibfnamefont {J.}~\bibnamefont {Verbeeck}}, \bibinfo {author} {\bibfnamefont {J.}~\bibnamefont {Manca}}, \bibinfo {author} {\bibfnamefont {E.}~\bibnamefont {Mosconi}}, \bibinfo {author} {\bibfnamefont {F.}~\bibnamefont {De~Angelis}},\ and\ \bibinfo {author} {\bibfnamefont {H.~G.}\ \bibnamefont {Boyen}},\ }\bibfield  {title} {\bibinfo {title} {\singleletter{Intrinsic Thermal Instability of Methylammonium Lead Trihalide Perovskite}},\ }\href {https://doi.org/10.1002/aenm.201500477} {\bibfield  {journal} {\bibinfo
  {journal} {Advanced Energy Materials}\ }\textbf {\bibinfo {volume} {5}},\ \bibinfo {pages} {1500477} (\bibinfo {year} {2015})}\BibitemShut {NoStop}%
\bibitem [{\citenamefont {Leijtens}\ \emph {et~al.}(2015)\citenamefont {Leijtens}, \citenamefont {Eperon}, \citenamefont {Noel}, \citenamefont {Habisreutinger}, \citenamefont {Petrozza},\ and\ \citenamefont {Snaith}}]{leijtens2015a}%
  \BibitemOpen
  \bibfield  {author} {\bibinfo {author} {\bibfnamefont {T.}~\bibnamefont {Leijtens}}, \bibinfo {author} {\bibfnamefont {G.~E.}\ \bibnamefont {Eperon}}, \bibinfo {author} {\bibfnamefont {N.~K.}\ \bibnamefont {Noel}}, \bibinfo {author} {\bibfnamefont {S.~N.}\ \bibnamefont {Habisreutinger}}, \bibinfo {author} {\bibfnamefont {A.}~\bibnamefont {Petrozza}},\ and\ \bibinfo {author} {\bibfnamefont {H.~J.}\ \bibnamefont {Snaith}},\ }\bibfield  {title} {\bibinfo {title} {\singleletter{Stability of metal halide perovskite solar cells}},\ }\href {https://doi.org/10.1002/aenm.201500963} {\bibfield  {journal} {\bibinfo  {journal} {Advanced Energy Materials}\ }\textbf {\bibinfo {volume} {5}},\ \bibinfo {pages} {1500963} (\bibinfo {year} {2015})}\BibitemShut {NoStop}%
\bibitem [{\citenamefont {Kim}\ \emph {et~al.}(2020)\citenamefont {Kim}, \citenamefont {Márquez}, \citenamefont {Unold},\ and\ \citenamefont {Walsh}}]{kim2020a}%
  \BibitemOpen
  \bibfield  {author} {\bibinfo {author} {\bibfnamefont {S.}~\bibnamefont {Kim}}, \bibinfo {author} {\bibfnamefont {J.~A.}\ \bibnamefont {Márquez}}, \bibinfo {author} {\bibfnamefont {T.}~\bibnamefont {Unold}},\ and\ \bibinfo {author} {\bibfnamefont {A.}~\bibnamefont {Walsh}},\ }\bibfield  {title} {\bibinfo {title} {\singleletter{Upper limit to the photovoltaic efficiency of imperfect crystals from first principles}},\ }\href {https://doi.org/10.1039/d0ee00291g} {\bibfield  {journal} {\bibinfo  {journal} {Energy and Environmental Science}\ }\textbf {\bibinfo {volume} {13}},\ \bibinfo {pages} {1481} (\bibinfo {year} {2020})}\BibitemShut {NoStop}%
\bibitem [{\citenamefont {Nielsen}\ \emph {et~al.}()\citenamefont {Nielsen}, \citenamefont {{\'A}lvarez}, \citenamefont {Tomm}, \citenamefont {Gurieva}, \citenamefont {Ortega-Guerrero}, \citenamefont {Breternitz}, \citenamefont {Bastonero}, \citenamefont {Marzari}, \citenamefont {Pignedoli}, \citenamefont {Schorr},\ and\ \citenamefont {Dimitrievska}}]{Nielsen2025a}%
  \BibitemOpen
  \bibfield  {author} {\bibinfo {author} {\bibfnamefont {R.~S.}\ \bibnamefont {Nielsen}}, \bibinfo {author} {\bibfnamefont {{\'A}.~L.}\ \bibnamefont {{\'A}lvarez}}, \bibinfo {author} {\bibfnamefont {Y.}~\bibnamefont {Tomm}}, \bibinfo {author} {\bibfnamefont {G.}~\bibnamefont {Gurieva}}, \bibinfo {author} {\bibfnamefont {A.}~\bibnamefont {Ortega-Guerrero}}, \bibinfo {author} {\bibfnamefont {J.}~\bibnamefont {Breternitz}}, \bibinfo {author} {\bibfnamefont {L.}~\bibnamefont {Bastonero}}, \bibinfo {author} {\bibfnamefont {N.}~\bibnamefont {Marzari}}, \bibinfo {author} {\bibfnamefont {C.~A.}\ \bibnamefont {Pignedoli}}, \bibinfo {author} {\bibfnamefont {S.}~\bibnamefont {Schorr}},\ and\ \bibinfo {author} {\bibfnamefont {M.}~\bibnamefont {Dimitrievska}},\ }\bibfield  {title} {\bibinfo {title} {\singleletter{BaZrS$_\text{3}$ Lights Up: The Interplay of Electrons, Photons, and Phonons in Strongly Luminescent Single Crystals}},\ }\href {https://doi.org/https://doi.org/10.1002/adom.202500915} {\bibfield  {journal}
  {\bibinfo  {journal} {Advanced Optical Materials}\ }\textbf {\bibinfo {volume} {n/a}},\ \bibinfo {pages} {e00915}}\BibitemShut {NoStop}%
\bibitem [{\citenamefont {Ghorpade}\ \emph {et~al.}(2023)\citenamefont {Ghorpade}, \citenamefont {Suryawanshi}, \citenamefont {Green}, \citenamefont {Wu}, \citenamefont {Hao},\ and\ \citenamefont {Ryan}}]{ghorpade2023a}%
  \BibitemOpen
  \bibfield  {author} {\bibinfo {author} {\bibfnamefont {U.~V.}\ \bibnamefont {Ghorpade}}, \bibinfo {author} {\bibfnamefont {M.~P.}\ \bibnamefont {Suryawanshi}}, \bibinfo {author} {\bibfnamefont {M.~A.}\ \bibnamefont {Green}}, \bibinfo {author} {\bibfnamefont {T.}~\bibnamefont {Wu}}, \bibinfo {author} {\bibfnamefont {X.}~\bibnamefont {Hao}},\ and\ \bibinfo {author} {\bibfnamefont {K.~M.}\ \bibnamefont {Ryan}},\ }\bibfield  {title} {\bibinfo {title} {\singleletter{Emerging Chalcohalide Materials for Energy Applications}},\ }\href {https://doi.org/10.1021/acs.chemrev.2c00422} {\bibfield  {journal} {\bibinfo  {journal} {Chemical Reviews}\ }\textbf {\bibinfo {volume} {123}},\ \bibinfo {pages} {327} (\bibinfo {year} {2023})}\BibitemShut {NoStop}%
\bibitem [{\citenamefont {Zhang}\ \emph {et~al.}(2025)\citenamefont {Zhang}, \citenamefont {Xia}, \citenamefont {Zhang}, \citenamefont {Ghorpade}, \citenamefont {He}, \citenamefont {Shin}, \citenamefont {Hao},\ and\ \citenamefont {Suryawanshi}}]{zhang2025a}%
  \BibitemOpen
  \bibfield  {author} {\bibinfo {author} {\bibfnamefont {H.}~\bibnamefont {Zhang}}, \bibinfo {author} {\bibfnamefont {Y.}~\bibnamefont {Xia}}, \bibinfo {author} {\bibfnamefont {Y.}~\bibnamefont {Zhang}}, \bibinfo {author} {\bibfnamefont {U.~V.}\ \bibnamefont {Ghorpade}}, \bibinfo {author} {\bibfnamefont {M.}~\bibnamefont {He}}, \bibinfo {author} {\bibfnamefont {S.~W.}\ \bibnamefont {Shin}}, \bibinfo {author} {\bibfnamefont {X.}~\bibnamefont {Hao}},\ and\ \bibinfo {author} {\bibfnamefont {M.~P.}\ \bibnamefont {Suryawanshi}},\ }\bibfield  {title} {\bibinfo {title} {\singleletter{The Rise of Chalcohalide Solar Cells: Comprehensive Insights From Materials to Devices}},\ }\href {https://doi.org/10.1002/advs.202413131} {\bibfield  {journal} {\bibinfo  {journal} {Advanced Science}\ ,\ \bibinfo {pages} {e2413131}} (\bibinfo {year} {2025})}\BibitemShut {NoStop}%
\bibitem [{\citenamefont {López}\ \emph {et~al.}(2024)\citenamefont {López}, \citenamefont {Caño}, \citenamefont {Rovira}, \citenamefont {Benítez}, \citenamefont {Asensi}, \citenamefont {Jehl}, \citenamefont {Tamarit}, \citenamefont {Saucedo},\ and\ \citenamefont {Cazorla}}]{l2024a}%
  \BibitemOpen
  \bibfield  {author} {\bibinfo {author} {\bibfnamefont {C.}~\bibnamefont {López}}, \bibinfo {author} {\bibfnamefont {I.}~\bibnamefont {Caño}}, \bibinfo {author} {\bibfnamefont {D.}~\bibnamefont {Rovira}}, \bibinfo {author} {\bibfnamefont {P.}~\bibnamefont {Benítez}}, \bibinfo {author} {\bibfnamefont {J.~M.}\ \bibnamefont {Asensi}}, \bibinfo {author} {\bibfnamefont {Z.}~\bibnamefont {Jehl}}, \bibinfo {author} {\bibfnamefont {J.~L.}\ \bibnamefont {Tamarit}}, \bibinfo {author} {\bibfnamefont {E.}~\bibnamefont {Saucedo}},\ and\ \bibinfo {author} {\bibfnamefont {C.}~\bibnamefont {Cazorla}},\ }\bibfield  {title} {\bibinfo {title} {\singleletter{Machine-Learning Aided First-Principles Prediction of Earth-Abundant Pnictogen Chalcohalide Solid Solutions for Solar-Cell Devices}},\ }\href {https://doi.org/10.1002/adfm.202406678} {\bibfield  {journal} {\bibinfo  {journal} {Advanced Functional Materials}\ }\textbf {\bibinfo {volume} {34}},\ \bibinfo {pages} {2406678} (\bibinfo {year} {2024})}\BibitemShut {NoStop}%
\bibitem [{\citenamefont {Guo}\ \emph {et~al.}(2023)\citenamefont {Guo}, \citenamefont {Huang}, \citenamefont {Lohan}, \citenamefont {Ye}, \citenamefont {Lin}, \citenamefont {Lim}, \citenamefont {Gauriot}, \citenamefont {Zelewski}, \citenamefont {Darvill}, \citenamefont {Zhu}, \citenamefont {Rao}, \citenamefont {McCulloch},\ and\ \citenamefont {Hoye}}]{guo2023a}%
  \BibitemOpen
  \bibfield  {author} {\bibinfo {author} {\bibfnamefont {X.}~\bibnamefont {Guo}}, \bibinfo {author} {\bibfnamefont {Y.~T.}\ \bibnamefont {Huang}}, \bibinfo {author} {\bibfnamefont {H.}~\bibnamefont {Lohan}}, \bibinfo {author} {\bibfnamefont {J.}~\bibnamefont {Ye}}, \bibinfo {author} {\bibfnamefont {Y.}~\bibnamefont {Lin}}, \bibinfo {author} {\bibfnamefont {J.}~\bibnamefont {Lim}}, \bibinfo {author} {\bibfnamefont {N.}~\bibnamefont {Gauriot}}, \bibinfo {author} {\bibfnamefont {S.~J.}\ \bibnamefont {Zelewski}}, \bibinfo {author} {\bibfnamefont {D.}~\bibnamefont {Darvill}}, \bibinfo {author} {\bibfnamefont {H.}~\bibnamefont {Zhu}}, \bibinfo {author} {\bibfnamefont {A.}~\bibnamefont {Rao}}, \bibinfo {author} {\bibfnamefont {I.}~\bibnamefont {McCulloch}},\ and\ \bibinfo {author} {\bibfnamefont {R.~L.}\ \bibnamefont {Hoye}},\ }\bibfield  {title} {\bibinfo {title} {\singleletter{Air-stable bismuth sulfobromide (BiSBr) visible-light absorbers: optoelectronic properties and potential for energy harvesting}},\ }\href
  {https://doi.org/10.1039/d3ta04491b} {\bibfield  {journal} {\bibinfo  {journal} {Journal of Materials Chemistry A}\ }\textbf {\bibinfo {volume} {11}},\ \bibinfo {pages} {22775} (\bibinfo {year} {2023})}\BibitemShut {NoStop}%
\bibitem [{\citenamefont {Palazon}(2022)}]{palazon2022a}%
  \BibitemOpen
  \bibfield  {author} {\bibinfo {author} {\bibfnamefont {F.}~\bibnamefont {Palazon}},\ }\bibfield  {title} {\bibinfo {title} {\singleletter{Metal Chalcohalides: Next Generation Photovoltaic Materials?}},\ }\href {https://doi.org/10.1002/solr.202100829} {\bibfield  {journal} {\bibinfo  {journal} {Solar RRL}\ }\textbf {\bibinfo {volume} {6}},\ \bibinfo {pages} {2100829} (\bibinfo {year} {2022})}\BibitemShut {NoStop}%
\bibitem [{\citenamefont {Wlaźlak}\ \emph {et~al.}(2018)\citenamefont {Wlaźlak}, \citenamefont {Blachecki}, \citenamefont {Bisztyga-Szklarz}, \citenamefont {Klejna}, \citenamefont {Mazur}, \citenamefont {Mech}, \citenamefont {Pilarczyk}, \citenamefont {Przyczyna}, \citenamefont {Suchecki}, \citenamefont {Zawal},\ and\ \citenamefont {Szaciłowski}}]{wla2018a}%
  \BibitemOpen
  \bibfield  {author} {\bibinfo {author} {\bibfnamefont {E.}~\bibnamefont {Wlaźlak}}, \bibinfo {author} {\bibfnamefont {A.}~\bibnamefont {Blachecki}}, \bibinfo {author} {\bibfnamefont {M.}~\bibnamefont {Bisztyga-Szklarz}}, \bibinfo {author} {\bibfnamefont {S.}~\bibnamefont {Klejna}}, \bibinfo {author} {\bibfnamefont {T.}~\bibnamefont {Mazur}}, \bibinfo {author} {\bibfnamefont {K.}~\bibnamefont {Mech}}, \bibinfo {author} {\bibfnamefont {K.}~\bibnamefont {Pilarczyk}}, \bibinfo {author} {\bibfnamefont {D.}~\bibnamefont {Przyczyna}}, \bibinfo {author} {\bibfnamefont {M.}~\bibnamefont {Suchecki}}, \bibinfo {author} {\bibfnamefont {P.}~\bibnamefont {Zawal}},\ and\ \bibinfo {author} {\bibfnamefont {K.}~\bibnamefont {Szaciłowski}},\ }\bibfield  {title} {\bibinfo {title} {\singleletter{Heavy pnictogen chalcohalides: The synthesis, structure and properties of these rediscovered semiconductors}},\ }\href {https://doi.org/10.1039/c8cc05149f} {\bibfield  {journal} {\bibinfo  {journal} {Chemical Communications}\ }\textbf
  {\bibinfo {volume} {54}},\ \bibinfo {pages} {12133} (\bibinfo {year} {2018})}\BibitemShut {NoStop}%
\bibitem [{\citenamefont {Sun}\ \emph {et~al.}(2016)\citenamefont {Sun}, \citenamefont {Shi}, \citenamefont {Lian}, \citenamefont {Gao}, \citenamefont {Agiorgousis}, \citenamefont {Zhang},\ and\ \citenamefont {Zhang}}]{sun2016a}%
  \BibitemOpen
  \bibfield  {author} {\bibinfo {author} {\bibfnamefont {Y.-Y.}\ \bibnamefont {Sun}}, \bibinfo {author} {\bibfnamefont {J.}~\bibnamefont {Shi}}, \bibinfo {author} {\bibfnamefont {J.}~\bibnamefont {Lian}}, \bibinfo {author} {\bibfnamefont {W.}~\bibnamefont {Gao}}, \bibinfo {author} {\bibfnamefont {M.~L.}\ \bibnamefont {Agiorgousis}}, \bibinfo {author} {\bibfnamefont {P.}~\bibnamefont {Zhang}},\ and\ \bibinfo {author} {\bibfnamefont {S.}~\bibnamefont {Zhang}},\ }\bibfield  {title} {\bibinfo {title} {\singleletter{Discovering lead-free perovskite solar materials with a split-anion approach}},\ }\href {https://doi.org/10.1039/C5NR04310G} {\bibfield  {journal} {\bibinfo  {journal} {Nanoscale}\ }\textbf {\bibinfo {volume} {8}},\ \bibinfo {pages} {6284} (\bibinfo {year} {2016})}\BibitemShut {NoStop}%
\bibitem [{\citenamefont {Kavanagh}\ \emph {et~al.}(2021)\citenamefont {Kavanagh}, \citenamefont {Savory}, \citenamefont {Scanlon},\ and\ \citenamefont {Walsh}}]{kavanagh2021a}%
  \BibitemOpen
  \bibfield  {author} {\bibinfo {author} {\bibfnamefont {S.~R.}\ \bibnamefont {Kavanagh}}, \bibinfo {author} {\bibfnamefont {C.~N.}\ \bibnamefont {Savory}}, \bibinfo {author} {\bibfnamefont {D.~O.}\ \bibnamefont {Scanlon}},\ and\ \bibinfo {author} {\bibfnamefont {A.}~\bibnamefont {Walsh}},\ }\bibfield  {title} {\bibinfo {title} {\singleletter{Hidden spontaneous polarisation in the chalcohalide photovoltaic absorber Sn$_\text{2}$SbS$_\text{2}$I$_\text{3}$}},\ }\href {https://doi.org/10.1039/d1mh00764e} {\bibfield  {journal} {\bibinfo  {journal} {Materials Horizons}\ }\textbf {\bibinfo {volume} {8}},\ \bibinfo {pages} {2709} (\bibinfo {year} {2021})}\BibitemShut {NoStop}%
\bibitem [{\citenamefont {Liu}\ \emph {et~al.}(2021{\natexlab{b}})\citenamefont {Liu}, \citenamefont {Mi}, \citenamefont {Ji}, \citenamefont {Liu}, \citenamefont {Fu}, \citenamefont {Hu}, \citenamefont {Xia},\ and\ \citenamefont {Xiao}}]{liu2021b}%
  \BibitemOpen
  \bibfield  {author} {\bibinfo {author} {\bibfnamefont {Z.}~\bibnamefont {Liu}}, \bibinfo {author} {\bibfnamefont {R.}~\bibnamefont {Mi}}, \bibinfo {author} {\bibfnamefont {G.}~\bibnamefont {Ji}}, \bibinfo {author} {\bibfnamefont {Y.}~\bibnamefont {Liu}}, \bibinfo {author} {\bibfnamefont {P.}~\bibnamefont {Fu}}, \bibinfo {author} {\bibfnamefont {S.}~\bibnamefont {Hu}}, \bibinfo {author} {\bibfnamefont {B.}~\bibnamefont {Xia}},\ and\ \bibinfo {author} {\bibfnamefont {Z.}~\bibnamefont {Xiao}},\ }\bibfield  {title} {\bibinfo {title} {\singleletter{Bandgap engineering and thermodynamic stability of oxyhalide and chalcohalide antiperovskites}},\ }\href {https://doi.org/10.1016/j.ceramint.2021.08.159} {\bibfield  {journal} {\bibinfo  {journal} {Ceramics International}\ }\textbf {\bibinfo {volume} {47}},\ \bibinfo {pages} {32634} (\bibinfo {year} {2021}{\natexlab{b}})}\BibitemShut {NoStop}%
\bibitem [{\citenamefont {Roth}\ \emph {et~al.}(2024)\citenamefont {Roth}, \citenamefont {Porter}, \citenamefont {Horger}, \citenamefont {Ochoa-Romero}, \citenamefont {Guirado}, \citenamefont {Rossini},\ and\ \citenamefont {Vela}}]{roth2024a}%
  \BibitemOpen
  \bibfield  {author} {\bibinfo {author} {\bibfnamefont {A.~N.}\ \bibnamefont {Roth}}, \bibinfo {author} {\bibfnamefont {A.~P.}\ \bibnamefont {Porter}}, \bibinfo {author} {\bibfnamefont {S.}~\bibnamefont {Horger}}, \bibinfo {author} {\bibfnamefont {K.}~\bibnamefont {Ochoa-Romero}}, \bibinfo {author} {\bibfnamefont {G.}~\bibnamefont {Guirado}}, \bibinfo {author} {\bibfnamefont {A.~J.}\ \bibnamefont {Rossini}},\ and\ \bibinfo {author} {\bibfnamefont {J.}~\bibnamefont {Vela}},\ }\bibfield  {title} {\bibinfo {title} {\singleletter{Lead-Free Semiconductors: Phase-Evolution and Superior Stability of Multinary Tin Chalcohalides}},\ }\href {https://doi.org/10.1021/acs.chemmater.4c00209} {\bibfield  {journal} {\bibinfo  {journal} {Chemistry of Materials}\ }\textbf {\bibinfo {volume} {36}},\ \bibinfo {pages} {4542} (\bibinfo {year} {2024})}\BibitemShut {NoStop}%
\bibitem [{\citenamefont {Brandt}\ \emph {et~al.}(2017)\citenamefont {Brandt}, \citenamefont {Poindexter}, \citenamefont {Gorai}, \citenamefont {Kurchin}, \citenamefont {Hoye}, \citenamefont {Nienhaus}, \citenamefont {Wilson}, \citenamefont {Polizzotti}, \citenamefont {Sereika}, \citenamefont {{\v{Z}}altauskas}, \citenamefont {Lee}, \citenamefont {MacManus-Driscoll}, \citenamefont {Bawendi}, \citenamefont {Stevanović},\ and\ \citenamefont {Buonassisi}}]{brandt2017a}%
  \BibitemOpen
  \bibfield  {author} {\bibinfo {author} {\bibfnamefont {R.~E.}\ \bibnamefont {Brandt}}, \bibinfo {author} {\bibfnamefont {J.~R.}\ \bibnamefont {Poindexter}}, \bibinfo {author} {\bibfnamefont {P.}~\bibnamefont {Gorai}}, \bibinfo {author} {\bibfnamefont {R.~C.}\ \bibnamefont {Kurchin}}, \bibinfo {author} {\bibfnamefont {R.~L.~Z.}\ \bibnamefont {Hoye}}, \bibinfo {author} {\bibfnamefont {L.}~\bibnamefont {Nienhaus}}, \bibinfo {author} {\bibfnamefont {M.~W.~B.}\ \bibnamefont {Wilson}}, \bibinfo {author} {\bibfnamefont {J.~A.}\ \bibnamefont {Polizzotti}}, \bibinfo {author} {\bibfnamefont {R.}~\bibnamefont {Sereika}}, \bibinfo {author} {\bibfnamefont {R.}~\bibnamefont {{\v{Z}}altauskas}}, \bibinfo {author} {\bibfnamefont {L.~C.}\ \bibnamefont {Lee}}, \bibinfo {author} {\bibfnamefont {J.~L.}\ \bibnamefont {MacManus-Driscoll}}, \bibinfo {author} {\bibfnamefont {M.}~\bibnamefont {Bawendi}}, \bibinfo {author} {\bibfnamefont {V.}~\bibnamefont {Stevanović}},\ and\ \bibinfo {author} {\bibfnamefont {T.}~\bibnamefont
  {Buonassisi}},\ }\bibfield  {title} {\bibinfo {title} {\singleletter{Searching for “Defect-Tolerant” Photovoltaic Materials: Combined Theoretical and Experimental Screening}},\ }\href {https://doi.org/10.1021/acs.chemmater.6b05496} {\bibfield  {journal} {\bibinfo  {journal} {Chemistry of Materials}\ }\textbf {\bibinfo {volume} {29}},\ \bibinfo {pages} {4667} (\bibinfo {year} {2017})}\BibitemShut {NoStop}%
\bibitem [{\citenamefont {Kurchin}\ \emph {et~al.}(2018)\citenamefont {Kurchin}, \citenamefont {Gorai}, \citenamefont {Buonassisi},\ and\ \citenamefont {Stevanović}}]{kurchin2018a}%
  \BibitemOpen
  \bibfield  {author} {\bibinfo {author} {\bibfnamefont {R.~C.}\ \bibnamefont {Kurchin}}, \bibinfo {author} {\bibfnamefont {P.}~\bibnamefont {Gorai}}, \bibinfo {author} {\bibfnamefont {T.}~\bibnamefont {Buonassisi}},\ and\ \bibinfo {author} {\bibfnamefont {V.}~\bibnamefont {Stevanović}},\ }\bibfield  {title} {\bibinfo {title} {\singleletter{Structural and Chemical Features Giving Rise to Defect Tolerance of Binary Semiconductors}},\ }\href {https://doi.org/10.1021/acs.chemmater.8b01505} {\bibfield  {journal} {\bibinfo  {journal} {Chemistry of Materials}\ }\textbf {\bibinfo {volume} {30}},\ \bibinfo {pages} {5583} (\bibinfo {year} {2018})}\BibitemShut {NoStop}%
\bibitem [{\citenamefont {Nicolson}\ \emph {et~al.}(2023)\citenamefont {Nicolson}, \citenamefont {Breternitz}, \citenamefont {Kavanagh}, \citenamefont {Tomm}, \citenamefont {Morita}, \citenamefont {Squires}, \citenamefont {Tovar}, \citenamefont {Walsh}, \citenamefont {Schorr},\ and\ \citenamefont {Scanlon}}]{nicolson2023a}%
  \BibitemOpen
  \bibfield  {author} {\bibinfo {author} {\bibfnamefont {A.}~\bibnamefont {Nicolson}}, \bibinfo {author} {\bibfnamefont {J.}~\bibnamefont {Breternitz}}, \bibinfo {author} {\bibfnamefont {S.~R.}\ \bibnamefont {Kavanagh}}, \bibinfo {author} {\bibfnamefont {Y.}~\bibnamefont {Tomm}}, \bibinfo {author} {\bibfnamefont {K.}~\bibnamefont {Morita}}, \bibinfo {author} {\bibfnamefont {A.~G.}\ \bibnamefont {Squires}}, \bibinfo {author} {\bibfnamefont {M.}~\bibnamefont {Tovar}}, \bibinfo {author} {\bibfnamefont {A.}~\bibnamefont {Walsh}}, \bibinfo {author} {\bibfnamefont {S.}~\bibnamefont {Schorr}},\ and\ \bibinfo {author} {\bibfnamefont {D.~O.}\ \bibnamefont {Scanlon}},\ }\bibfield  {title} {\bibinfo {title} {\singleletter{Interplay of Static and Dynamic Disorder in the Mixed-Metal Chalcohalide Sn$_\text{2}$SbS$_\text{2}$I$_\text{3}$}},\ }\href {https://doi.org/10.1021/jacs.2c13336} {\bibfield  {journal} {\bibinfo  {journal} {Journal of the American Chemical Society}\ }\textbf {\bibinfo {volume} {145}},\ \bibinfo {pages}
  {12509} (\bibinfo {year} {2023})}\BibitemShut {NoStop}%
\bibitem [{\citenamefont {Ran}\ \emph {et~al.}(2018)\citenamefont {Ran}, \citenamefont {Wang}, \citenamefont {Li}, \citenamefont {Yang}, \citenamefont {Zhao}, \citenamefont {Biswas}, \citenamefont {Singh},\ and\ \citenamefont {Zhang}}]{ran2018a}%
  \BibitemOpen
  \bibfield  {author} {\bibinfo {author} {\bibfnamefont {Z.}~\bibnamefont {Ran}}, \bibinfo {author} {\bibfnamefont {X.}~\bibnamefont {Wang}}, \bibinfo {author} {\bibfnamefont {Y.}~\bibnamefont {Li}}, \bibinfo {author} {\bibfnamefont {D.}~\bibnamefont {Yang}}, \bibinfo {author} {\bibfnamefont {X.~G.}\ \bibnamefont {Zhao}}, \bibinfo {author} {\bibfnamefont {K.}~\bibnamefont {Biswas}}, \bibinfo {author} {\bibfnamefont {D.~J.}\ \bibnamefont {Singh}},\ and\ \bibinfo {author} {\bibfnamefont {L.}~\bibnamefont {Zhang}},\ }\bibfield  {title} {\bibinfo {title} {\singleletter{Bismuth and antimony-based oxyhalides and chalcohalides as potential optoelectronic materials}},\ }\href {https://doi.org/10.1038/s41524-018-0071-1} {\bibfield  {journal} {\bibinfo  {journal} {npj Computational Materials}\ }\textbf {\bibinfo {volume} {4}},\ \bibinfo {pages} {14} (\bibinfo {year} {2018})}\BibitemShut {NoStop}%
\bibitem [{\citenamefont {Ganose}\ \emph {et~al.}(2016)\citenamefont {Ganose}, \citenamefont {Butler}, \citenamefont {Walsh},\ and\ \citenamefont {Scanlon}}]{ganose2016a}%
  \BibitemOpen
  \bibfield  {author} {\bibinfo {author} {\bibfnamefont {A.~M.}\ \bibnamefont {Ganose}}, \bibinfo {author} {\bibfnamefont {K.~T.}\ \bibnamefont {Butler}}, \bibinfo {author} {\bibfnamefont {A.}~\bibnamefont {Walsh}},\ and\ \bibinfo {author} {\bibfnamefont {D.~O.}\ \bibnamefont {Scanlon}},\ }\bibfield  {title} {\bibinfo {title} {\singleletter{Relativistic electronic structure and band alignment of BiSI and BiSeI: candidate photovoltaic materials}},\ }\href {https://doi.org/10.1039/c5ta09612j} {\bibfield  {journal} {\bibinfo  {journal} {Journal of Materials Chemistry A}\ }\textbf {\bibinfo {volume} {4}},\ \bibinfo {pages} {2060} (\bibinfo {year} {2016})}\BibitemShut {NoStop}%
\bibitem [{\citenamefont {Zunger}(2018)}]{zunger2018a}%
  \BibitemOpen
  \bibfield  {author} {\bibinfo {author} {\bibfnamefont {A.}~\bibnamefont {Zunger}},\ }\bibfield  {title} {\bibinfo {title} {\singleletter{Inverse design in search of materials with target functionalities}},\ }\href {https://doi.org/10.1038/S41570-018-0121} {\bibfield  {journal} {\bibinfo  {journal} {Nature Reviews Chemistry}\ }\textbf {\bibinfo {volume} {2}},\ \bibinfo {pages} {0121} (\bibinfo {year} {2018})}\BibitemShut {NoStop}%
\bibitem [{\citenamefont {Caño}\ \emph {et~al.}(2025)\citenamefont {Caño}, \citenamefont {Navarro-Güell}, \citenamefont {Maggi}, \citenamefont {Gon~Medaille}, \citenamefont {Rovira}, \citenamefont {Jimenez-Arguijo}, \citenamefont {Segura}, \citenamefont {Torrens}, \citenamefont {Jimenez}, \citenamefont {López}, \citenamefont {Benítez}, \citenamefont {Cazorla}, \citenamefont {Jehl}, \citenamefont {Gong}, \citenamefont {Asensi}, \citenamefont {Calvo-Barrio}, \citenamefont {Soler}, \citenamefont {Llorca}, \citenamefont {Tamarit}, \citenamefont {Galiana}, \citenamefont {Dimitrievska}, \citenamefont {Ruiz-Marín}, \citenamefont {Chun}, \citenamefont {Wong}, \citenamefont {Puigdollers}, \citenamefont {Placidi},\ and\ \citenamefont {Saucedo}}]{ca2025a}%
  \BibitemOpen
  \bibfield  {author} {\bibinfo {author} {\bibfnamefont {I.}~\bibnamefont {Caño}}, \bibinfo {author} {\bibfnamefont {A.}~\bibnamefont {Navarro-Güell}}, \bibinfo {author} {\bibfnamefont {E.}~\bibnamefont {Maggi}}, \bibinfo {author} {\bibfnamefont {A.}~\bibnamefont {Gon~Medaille}}, \bibinfo {author} {\bibfnamefont {D.}~\bibnamefont {Rovira}}, \bibinfo {author} {\bibfnamefont {A.}~\bibnamefont {Jimenez-Arguijo}}, \bibinfo {author} {\bibfnamefont {O.}~\bibnamefont {Segura}}, \bibinfo {author} {\bibfnamefont {A.}~\bibnamefont {Torrens}}, \bibinfo {author} {\bibfnamefont {M.}~\bibnamefont {Jimenez}}, \bibinfo {author} {\bibfnamefont {C.}~\bibnamefont {López}}, \bibinfo {author} {\bibfnamefont {P.}~\bibnamefont {Benítez}}, \bibinfo {author} {\bibfnamefont {C.}~\bibnamefont {Cazorla}}, \bibinfo {author} {\bibfnamefont {Z.}~\bibnamefont {Jehl}}, \bibinfo {author} {\bibfnamefont {Y.}~\bibnamefont {Gong}}, \bibinfo {author} {\bibfnamefont {J.-M.}\ \bibnamefont {Asensi}}, \bibinfo {author} {\bibfnamefont
  {L.}~\bibnamefont {Calvo-Barrio}}, \bibinfo {author} {\bibfnamefont {L.}~\bibnamefont {Soler}}, \bibinfo {author} {\bibfnamefont {J.}~\bibnamefont {Llorca}}, \bibinfo {author} {\bibfnamefont {J.-L.}\ \bibnamefont {Tamarit}}, \bibinfo {author} {\bibfnamefont {B.}~\bibnamefont {Galiana}}, \bibinfo {author} {\bibfnamefont {M.}~\bibnamefont {Dimitrievska}}, \bibinfo {author} {\bibfnamefont {N.}~\bibnamefont {Ruiz-Marín}}, \bibinfo {author} {\bibfnamefont {H.~Z.}\ \bibnamefont {Chun}}, \bibinfo {author} {\bibfnamefont {L.}~\bibnamefont {Wong}}, \bibinfo {author} {\bibfnamefont {J.}~\bibnamefont {Puigdollers}}, \bibinfo {author} {\bibfnamefont {M.}~\bibnamefont {Placidi}},\ and\ \bibinfo {author} {\bibfnamefont {E.}~\bibnamefont {Saucedo}},\ }\bibfield  {title} {\bibinfo {title} {\singleletter{Ribbons of Light: Emerging (Sb,Bi)(S,Se)(Br,I) Van der Waals Chalcohalides for Next-Generation Energy Applications}},\ }\href {https://doi.org/10.1002/smll.202505430} {\bibfield  {journal} {\bibinfo  {journal} {Small}\ ,\
  \bibinfo {pages} {e05430}} (\bibinfo {year} {2025})}\BibitemShut {NoStop}%
\bibitem [{\citenamefont {Kohn}\ and\ \citenamefont {Sham}(1965)}]{kohn1965a}%
  \BibitemOpen
  \bibfield  {author} {\bibinfo {author} {\bibfnamefont {W.}~\bibnamefont {Kohn}}\ and\ \bibinfo {author} {\bibfnamefont {L.~J.}\ \bibnamefont {Sham}},\ }\bibfield  {title} {\bibinfo {title} {\singleletter{Self-Consistent Equations Including Exchange and Correlation Effects}},\ }\href {https://doi.org/10.1103/PhysRev.140.A1133} {\bibfield  {journal} {\bibinfo  {journal} {Physical Review}\ }\textbf {\bibinfo {volume} {140}},\ \bibinfo {pages} {A1133} (\bibinfo {year} {1965})}\BibitemShut {NoStop}%
\bibitem [{\citenamefont {Kresse}\ and\ \citenamefont {Furthmüller}(1996)}]{kresse1996a}%
  \BibitemOpen
  \bibfield  {author} {\bibinfo {author} {\bibfnamefont {G.}~\bibnamefont {Kresse}}\ and\ \bibinfo {author} {\bibfnamefont {J.}~\bibnamefont {Furthmüller}},\ }\bibfield  {title} {\bibinfo {title} {\singleletter{Efficient iterative schemes for \textit{ab initio} total-energy calculations using a plane-wave basis set}},\ }\href {https://doi.org/10.1103/PhysRevB.54.11169} {\bibfield  {journal} {\bibinfo  {journal} {Physical Review B}\ }\textbf {\bibinfo {volume} {54}},\ \bibinfo {pages} {11169} (\bibinfo {year} {1996})}\BibitemShut {NoStop}%
\bibitem [{\citenamefont {Kresse}(1999)}]{kresse1999a}%
  \BibitemOpen
  \bibfield  {author} {\bibinfo {author} {\bibfnamefont {G.}~\bibnamefont {Kresse}},\ }\bibfield  {title} {\bibinfo {title} {\singleletter{From ultrasoft pseudopotentials to the projector augmented-wave method}},\ }\href {https://doi.org/10.1103/PhysRevB.59.1758} {\bibfield  {journal} {\bibinfo  {journal} {Physical Review B}\ }\textbf {\bibinfo {volume} {59}},\ \bibinfo {pages} {1758} (\bibinfo {year} {1999})}\BibitemShut {NoStop}%
\bibitem [{\citenamefont {Blöchl}(1994)}]{blochl1994a}%
  \BibitemOpen
  \bibfield  {author} {\bibinfo {author} {\bibfnamefont {P.~E.}\ \bibnamefont {Blöchl}},\ }\bibfield  {title} {\bibinfo {title} {\singleletter{Projector augmented-wave method}},\ }\href {https://doi.org/10.1103/PhysRevB.50.17953} {\bibfield  {journal} {\bibinfo  {journal} {Physical Review B}\ }\textbf {\bibinfo {volume} {50}},\ \bibinfo {pages} {17953} (\bibinfo {year} {1994})}\BibitemShut {NoStop}%
\bibitem [{\citenamefont {Perdew}\ \emph {et~al.}(2008)\citenamefont {Perdew}, \citenamefont {Ruzsinszky}, \citenamefont {Csonka}, \citenamefont {Vydrov}, \citenamefont {Scuseria}, \citenamefont {Constantin}, \citenamefont {Zhou},\ and\ \citenamefont {Burke}}]{perdew2008a}%
  \BibitemOpen
  \bibfield  {author} {\bibinfo {author} {\bibfnamefont {J.~P.}\ \bibnamefont {Perdew}}, \bibinfo {author} {\bibfnamefont {A.}~\bibnamefont {Ruzsinszky}}, \bibinfo {author} {\bibfnamefont {G.~I.}\ \bibnamefont {Csonka}}, \bibinfo {author} {\bibfnamefont {O.~A.}\ \bibnamefont {Vydrov}}, \bibinfo {author} {\bibfnamefont {G.~E.}\ \bibnamefont {Scuseria}}, \bibinfo {author} {\bibfnamefont {L.~A.}\ \bibnamefont {Constantin}}, \bibinfo {author} {\bibfnamefont {X.}~\bibnamefont {Zhou}},\ and\ \bibinfo {author} {\bibfnamefont {K.}~\bibnamefont {Burke}},\ }\bibfield  {title} {\bibinfo {title} {\singleletter{Restoring the Density-Gradient Expansion for Exchange in Solids and Surfaces}},\ }\href {https://doi.org/10.1103/PhysRevLett.100.136406} {\bibfield  {journal} {\bibinfo  {journal} {Physical Review Letters}\ }\textbf {\bibinfo {volume} {100}},\ \bibinfo {pages} {136406} (\bibinfo {year} {2008})}\BibitemShut {NoStop}%
\bibitem [{\citenamefont {Heyd}\ \emph {et~al.}(2003)\citenamefont {Heyd}, \citenamefont {Scuseria},\ and\ \citenamefont {Ernzerhof}}]{heyd2003a}%
  \BibitemOpen
  \bibfield  {author} {\bibinfo {author} {\bibfnamefont {J.}~\bibnamefont {Heyd}}, \bibinfo {author} {\bibfnamefont {G.~E.}\ \bibnamefont {Scuseria}},\ and\ \bibinfo {author} {\bibfnamefont {M.}~\bibnamefont {Ernzerhof}},\ }\bibfield  {title} {\bibinfo {title} {\singleletter{Hybrid functionals based on a screened Coulomb potential}},\ }\href {https://doi.org/10.1063/1.1564060} {\bibfield  {journal} {\bibinfo  {journal} {Journal of Chemical Physics}\ }\textbf {\bibinfo {volume} {118}},\ \bibinfo {pages} {8207} (\bibinfo {year} {2003})}\BibitemShut {NoStop}%
\bibitem [{\citenamefont {Schimka}\ \emph {et~al.}(2011)\citenamefont {Schimka}, \citenamefont {Harl},\ and\ \citenamefont {Kresse}}]{schimka2011a}%
  \BibitemOpen
  \bibfield  {author} {\bibinfo {author} {\bibfnamefont {L.}~\bibnamefont {Schimka}}, \bibinfo {author} {\bibfnamefont {J.}~\bibnamefont {Harl}},\ and\ \bibinfo {author} {\bibfnamefont {G.}~\bibnamefont {Kresse}},\ }\bibfield  {title} {\bibinfo {title} {\singleletter{Improved hybrid functional for solids: The HSEsol functional}},\ }\href {https://doi.org/10.1063/1.3524336} {\bibfield  {journal} {\bibinfo  {journal} {Journal of Chemical Physics}\ }\textbf {\bibinfo {volume} {134}},\ \bibinfo {pages} {024116} (\bibinfo {year} {2011})}\BibitemShut {NoStop}%
\bibitem [{\citenamefont {Grimme}\ \emph {et~al.}(2010)\citenamefont {Grimme}, \citenamefont {Antony}, \citenamefont {Ehrlich},\ and\ \citenamefont {Krieg}}]{grimme2010a}%
  \BibitemOpen
  \bibfield  {author} {\bibinfo {author} {\bibfnamefont {S.}~\bibnamefont {Grimme}}, \bibinfo {author} {\bibfnamefont {J.}~\bibnamefont {Antony}}, \bibinfo {author} {\bibfnamefont {S.}~\bibnamefont {Ehrlich}},\ and\ \bibinfo {author} {\bibfnamefont {H.}~\bibnamefont {Krieg}},\ }\bibfield  {title} {\bibinfo {title} {\singleletter{A consistent and accurate \textit{ab initio} parametrization of density functional dispersion correction (DFT-D) for the 94 elements H-Pu}},\ }\href {https://doi.org/10.1063/1.3382344} {\bibfield  {journal} {\bibinfo  {journal} {Journal of Chemical Physics}\ }\textbf {\bibinfo {volume} {132}},\ \bibinfo {pages} {154104} (\bibinfo {year} {2010})}\BibitemShut {NoStop}%
\bibitem [{\citenamefont {Togo}\ and\ \citenamefont {Tanaka}(2015)}]{togo2015a}%
  \BibitemOpen
  \bibfield  {author} {\bibinfo {author} {\bibfnamefont {A.}~\bibnamefont {Togo}}\ and\ \bibinfo {author} {\bibfnamefont {I.}~\bibnamefont {Tanaka}},\ }\bibfield  {title} {\bibinfo {title} {\singleletter{First principles phonon calculations in materials science}},\ }\href {https://doi.org/10.1016/j.scriptamat.2015.07.021} {\bibfield  {journal} {\bibinfo  {journal} {Scripta Materialia}\ }\textbf {\bibinfo {volume} {108}},\ \bibinfo {pages} {1} (\bibinfo {year} {2015})}\BibitemShut {NoStop}%
\bibitem [{\citenamefont {Horak}\ and\ \citenamefont {Cermak}(1965)}]{horak1965a}%
  \BibitemOpen
  \bibfield  {author} {\bibinfo {author} {\bibfnamefont {J.}~\bibnamefont {Horak}}\ and\ \bibinfo {author} {\bibfnamefont {K.}~\bibnamefont {Cermak}},\ }\bibfield  {title} {\bibinfo {title} {\singleletter{Preparation and photoelectric properties of bismuth sulphidiodide}},\ }\href {https://doi.org/10.1007/BF01689287} {\bibfield  {journal} {\bibinfo  {journal} {Czechoslovak Journal of Physics}\ }\textbf {\bibinfo {volume} {15}},\ \bibinfo {pages} {536} (\bibinfo {year} {1965})}\BibitemShut {NoStop}%
\bibitem [{\citenamefont {Sasaki}(1965)}]{sasaki1965a}%
  \BibitemOpen
  \bibfield  {author} {\bibinfo {author} {\bibfnamefont {Y.}~\bibnamefont {Sasaki}},\ }\bibfield  {title} {\bibinfo {title} {\singleletter{Photoconductivity of a Ferroelectric Photoconductor BiSI}},\ }\href {https://doi.org/10.1143/JJAP.4.614} {\bibfield  {journal} {\bibinfo  {journal} {Japanese Journal of Applied Physics}\ }\textbf {\bibinfo {volume} {4}},\ \bibinfo {pages} {614} (\bibinfo {year} {1965})}\BibitemShut {NoStop}%
\bibitem [{\citenamefont {Fa}\ \emph {et~al.}(2011)\citenamefont {Fa}, \citenamefont {Li}, \citenamefont {Zhang}, \citenamefont {Guo}, \citenamefont {Guo},\ and\ \citenamefont {Yang}}]{fa2011a}%
  \BibitemOpen
  \bibfield  {author} {\bibinfo {author} {\bibfnamefont {W.~J.}\ \bibnamefont {Fa}}, \bibinfo {author} {\bibfnamefont {P.~J.}\ \bibnamefont {Li}}, \bibinfo {author} {\bibfnamefont {Y.~G.}\ \bibnamefont {Zhang}}, \bibinfo {author} {\bibfnamefont {L.~L.}\ \bibnamefont {Guo}}, \bibinfo {author} {\bibfnamefont {J.~F.}\ \bibnamefont {Guo}},\ and\ \bibinfo {author} {\bibfnamefont {F.~L.}\ \bibnamefont {Yang}},\ }\bibfield  {title} {\bibinfo {title} {\singleletter{The competitive growth of BiOI and BiSI in the solvothermal process}},\ }\href {https://doi.org/10.4028/www.scientific.net/AMR.236-238.1919} {\bibfield  {journal} {\bibinfo  {journal} {Advanced Materials Research}\ }\textbf {\bibinfo {volume} {236-238}},\ \bibinfo {pages} {1919} (\bibinfo {year} {2011})}\BibitemShut {NoStop}%
\bibitem [{\citenamefont {Su}\ \emph {et~al.}(2006)\citenamefont {Su}, \citenamefont {Zhang}, \citenamefont {Liu}, \citenamefont {Liu}, \citenamefont {Qin},\ and\ \citenamefont {Chen}}]{su2006a}%
  \BibitemOpen
  \bibfield  {author} {\bibinfo {author} {\bibfnamefont {X.}~\bibnamefont {Su}}, \bibinfo {author} {\bibfnamefont {G.}~\bibnamefont {Zhang}}, \bibinfo {author} {\bibfnamefont {T.}~\bibnamefont {Liu}}, \bibinfo {author} {\bibfnamefont {Y.}~\bibnamefont {Liu}}, \bibinfo {author} {\bibfnamefont {J.}~\bibnamefont {Qin}},\ and\ \bibinfo {author} {\bibfnamefont {C.}~\bibnamefont {Chen}},\ }\bibfield  {title} {\bibinfo {title} {\singleletter{A facile and clean synthesis of pure bismuth sulfide iodide crystals}},\ }\href {https://doi.org/10.1134/S0036023606120047} {\bibfield  {journal} {\bibinfo  {journal} {Russian Journal of Inorganic Chemistry}\ }\textbf {\bibinfo {volume} {51}},\ \bibinfo {pages} {1864} (\bibinfo {year} {2006})}\BibitemShut {NoStop}%
\bibitem [{\citenamefont {Ganesha}\ \emph {et~al.}(1993)\citenamefont {Ganesha}, \citenamefont {Arivuoli},\ and\ \citenamefont {Ramasamy}}]{ganesha1993a}%
  \BibitemOpen
  \bibfield  {author} {\bibinfo {author} {\bibfnamefont {R.}~\bibnamefont {Ganesha}}, \bibinfo {author} {\bibfnamefont {D.}~\bibnamefont {Arivuoli}},\ and\ \bibinfo {author} {\bibfnamefont {P.}~\bibnamefont {Ramasamy}},\ }\bibfield  {title} {\bibinfo {title} {\singleletter{Growth of some group V-VI-VII compounds from the vapour}},\ }\href {https://doi.org/10.1016/S0022-0248(07)80101-1} {\bibfield  {journal} {\bibinfo  {journal} {Journal of Crystal Growth}\ }\textbf {\bibinfo {volume} {128}},\ \bibinfo {pages} {1081} (\bibinfo {year} {1993})}\BibitemShut {NoStop}%
\bibitem [{\citenamefont {Zhu}\ \emph {et~al.}(2003)\citenamefont {Zhu}, \citenamefont {Zheng}, \citenamefont {Yin}, \citenamefont {Liu}, \citenamefont {Jia},\ and\ \citenamefont {Xie}}]{zhu2003a}%
  \BibitemOpen
  \bibfield  {author} {\bibinfo {author} {\bibfnamefont {L.}~\bibnamefont {Zhu}}, \bibinfo {author} {\bibfnamefont {X.}~\bibnamefont {Zheng}}, \bibinfo {author} {\bibfnamefont {X.}~\bibnamefont {Yin}}, \bibinfo {author} {\bibfnamefont {X.}~\bibnamefont {Liu}}, \bibinfo {author} {\bibfnamefont {Y.}~\bibnamefont {Jia}},\ and\ \bibinfo {author} {\bibfnamefont {Y.}~\bibnamefont {Xie}},\ }\bibfield  {title} {\bibinfo {title} {\singleletter{A Mild Solution Route to Bismuth Selenoiodide Rod-like Crystals}},\ }\href {https://doi.org/10.1246/cl.2003.350} {\bibfield  {journal} {\bibinfo  {journal} {Chemistry Letters}\ }\textbf {\bibinfo {volume} {32}},\ \bibinfo {pages} {350} (\bibinfo {year} {2003})}\BibitemShut {NoStop}%
\bibitem [{\citenamefont {Choi}\ and\ \citenamefont {Hwang}(2019)}]{choi2019a}%
  \BibitemOpen
  \bibfield  {author} {\bibinfo {author} {\bibfnamefont {Y.~C.}\ \bibnamefont {Choi}}\ and\ \bibinfo {author} {\bibfnamefont {E.}~\bibnamefont {Hwang}},\ }\bibfield  {title} {\bibinfo {title} {\singleletter{Controlled Growth of BiSI Nanorod-Based Films through a Two-Step Solution Process for Solar Cell Applications}},\ }\href {https://doi.org/10.3390/nano9121650} {\bibfield  {journal} {\bibinfo  {journal} {Nanomaterials}\ }\textbf {\bibinfo {volume} {9}},\ \bibinfo {pages} {1650} (\bibinfo {year} {2019})}\BibitemShut {NoStop}%
\bibitem [{\citenamefont {Tiwari}\ \emph {et~al.}(2019)\citenamefont {Tiwari}, \citenamefont {Cardoso-Delgado}, \citenamefont {Alibhai}, \citenamefont {Mombrú},\ and\ \citenamefont {Fermín}}]{tiwari2019a}%
  \BibitemOpen
  \bibfield  {author} {\bibinfo {author} {\bibfnamefont {D.}~\bibnamefont {Tiwari}}, \bibinfo {author} {\bibfnamefont {F.}~\bibnamefont {Cardoso-Delgado}}, \bibinfo {author} {\bibfnamefont {D.}~\bibnamefont {Alibhai}}, \bibinfo {author} {\bibfnamefont {M.}~\bibnamefont {Mombrú}},\ and\ \bibinfo {author} {\bibfnamefont {D.~J.}\ \bibnamefont {Fermín}},\ }\bibfield  {title} {\bibinfo {title} {\singleletter{Photovoltaic Performance of Phase-Pure Orthorhombic BiSI Thin-Films}},\ }\href {https://doi.org/10.1021/acsaem.9b00544} {\bibfield  {journal} {\bibinfo  {journal} {ACS Applied Energy Materials}\ }\textbf {\bibinfo {volume} {2}},\ \bibinfo {pages} {3878} (\bibinfo {year} {2019})}\BibitemShut {NoStop}%
\bibitem [{\citenamefont {Kunioku}\ \emph {et~al.}(2016)\citenamefont {Kunioku}, \citenamefont {Higashi},\ and\ \citenamefont {Abe}}]{kunioku2016a}%
  \BibitemOpen
  \bibfield  {author} {\bibinfo {author} {\bibfnamefont {H.}~\bibnamefont {Kunioku}}, \bibinfo {author} {\bibfnamefont {M.}~\bibnamefont {Higashi}},\ and\ \bibinfo {author} {\bibfnamefont {R.}~\bibnamefont {Abe}},\ }\bibfield  {title} {\bibinfo {title} {\singleletter{Lowerature Synthesis of Bismuth Chalcohalides: Candidate Photovoltaic Materials with Easily, Continuously Controllable Band gap}},\ }\href {https://doi.org/10.1038/srep32664} {\bibfield  {journal} {\bibinfo  {journal} {Scientific Reports}\ }\textbf {\bibinfo {volume} {6}},\ \bibinfo {pages} {32664} (\bibinfo {year} {2016})}\BibitemShut {NoStop}%
\bibitem [{\citenamefont {Hahn}\ \emph {et~al.}(2012)\citenamefont {Hahn}, \citenamefont {Self},\ and\ \citenamefont {Mullins}}]{hahn2012a}%
  \BibitemOpen
  \bibfield  {author} {\bibinfo {author} {\bibfnamefont {N.~T.}\ \bibnamefont {Hahn}}, \bibinfo {author} {\bibfnamefont {J.~L.}\ \bibnamefont {Self}},\ and\ \bibinfo {author} {\bibfnamefont {C.~B.}\ \bibnamefont {Mullins}},\ }\bibfield  {title} {\bibinfo {title} {\singleletter{BiSI Micro-Rod Thin Films: Efficient Solar Absorber Electrodes?}},\ }\href {https://doi.org/10.1021/jz300515p} {\bibfield  {journal} {\bibinfo  {journal} {Journal of Physical Chemistry Letters}\ }\textbf {\bibinfo {volume} {3}},\ \bibinfo {pages} {1571} (\bibinfo {year} {2012})}\BibitemShut {NoStop}%
\bibitem [{\citenamefont {Zhou}\ \emph {et~al.}(2019)\citenamefont {Zhou}, \citenamefont {Wang}, \citenamefont {Jiang}, \citenamefont {Chen},\ and\ \citenamefont {Wang}}]{zhou2019a}%
  \BibitemOpen
  \bibfield  {author} {\bibinfo {author} {\bibfnamefont {C.}~\bibnamefont {Zhou}}, \bibinfo {author} {\bibfnamefont {R.}~\bibnamefont {Wang}}, \bibinfo {author} {\bibfnamefont {C.}~\bibnamefont {Jiang}}, \bibinfo {author} {\bibfnamefont {J.}~\bibnamefont {Chen}},\ and\ \bibinfo {author} {\bibfnamefont {G.}~\bibnamefont {Wang}},\ }\bibfield  {title} {\bibinfo {title} {\singleletter{Dynamically Optimized Multi-interface Novel BiSI-Promoted Redox Sites Spatially Separated n-p-n Double Heterojunctions BiSI/MoS$_\text{2}$/CdS for Hydrogen Evolution}},\ }\href {https://doi.org/10.1021/acs.iecr.9b00234} {\bibfield  {journal} {\bibinfo  {journal} {Industrial and Engineering Chemistry Research}\ }\textbf {\bibinfo {volume} {58}},\ \bibinfo {pages} {7844} (\bibinfo {year} {2019})}\BibitemShut {NoStop}%
\bibitem [{\citenamefont {Farooq}\ \emph {et~al.}(2021)\citenamefont {Farooq}, \citenamefont {Feeney}, \citenamefont {Mendes}, \citenamefont {Krishnamurthi}, \citenamefont {Walia}, \citenamefont {Della~Gaspera},\ and\ \citenamefont {van Embden}}]{farooq2021a}%
  \BibitemOpen
  \bibfield  {author} {\bibinfo {author} {\bibfnamefont {S.}~\bibnamefont {Farooq}}, \bibinfo {author} {\bibfnamefont {T.}~\bibnamefont {Feeney}}, \bibinfo {author} {\bibfnamefont {J.~O.}\ \bibnamefont {Mendes}}, \bibinfo {author} {\bibfnamefont {V.}~\bibnamefont {Krishnamurthi}}, \bibinfo {author} {\bibfnamefont {S.}~\bibnamefont {Walia}}, \bibinfo {author} {\bibfnamefont {E.}~\bibnamefont {Della~Gaspera}},\ and\ \bibinfo {author} {\bibfnamefont {J.}~\bibnamefont {van Embden}},\ }\bibfield  {title} {\bibinfo {title} {\singleletter{High Gain Solution-Processed Carbon-Free BiSI Chalcohalide Thin Film Photodetectors}},\ }\href {https://doi.org/10.1002/adfm.202104788} {\bibfield  {journal} {\bibinfo  {journal} {Advanced Functional Materials}\ }\textbf {\bibinfo {volume} {31}},\ \bibinfo {pages} {2104788} (\bibinfo {year} {2021})}\BibitemShut {NoStop}%
\bibitem [{\citenamefont {Jain}\ \emph {et~al.}(2013)\citenamefont {Jain}, \citenamefont {Ong}, \citenamefont {Hautier}, \citenamefont {Chen}, \citenamefont {Richards}, \citenamefont {Dacek}, \citenamefont {Cholia}, \citenamefont {Gunter}, \citenamefont {Skinner}, \citenamefont {Ceder},\ and\ \citenamefont {Persson}}]{jain2013a}%
  \BibitemOpen
  \bibfield  {author} {\bibinfo {author} {\bibfnamefont {A.}~\bibnamefont {Jain}}, \bibinfo {author} {\bibfnamefont {S.~P.}\ \bibnamefont {Ong}}, \bibinfo {author} {\bibfnamefont {G.}~\bibnamefont {Hautier}}, \bibinfo {author} {\bibfnamefont {W.}~\bibnamefont {Chen}}, \bibinfo {author} {\bibfnamefont {W.~D.}\ \bibnamefont {Richards}}, \bibinfo {author} {\bibfnamefont {S.}~\bibnamefont {Dacek}}, \bibinfo {author} {\bibfnamefont {S.}~\bibnamefont {Cholia}}, \bibinfo {author} {\bibfnamefont {D.}~\bibnamefont {Gunter}}, \bibinfo {author} {\bibfnamefont {D.}~\bibnamefont {Skinner}}, \bibinfo {author} {\bibfnamefont {G.}~\bibnamefont {Ceder}},\ and\ \bibinfo {author} {\bibfnamefont {K.~A.}\ \bibnamefont {Persson}},\ }\bibfield  {title} {\bibinfo {title} {\singleletter{Commentary: The Materials Project: A materials genome approach to accelerating materials innovation}},\ }\href {https://doi.org/10.1063/1.4812323} {\bibfield  {journal} {\bibinfo  {journal} {APL Materials}\ }\textbf {\bibinfo {volume} {1}},\ \bibinfo
  {pages} {011002} (\bibinfo {year} {2013})}\BibitemShut {NoStop}%
\bibitem [{\citenamefont {Nitsche}\ and\ \citenamefont {Merz}(1960)}]{nitsche1960a}%
  \BibitemOpen
  \bibfield  {author} {\bibinfo {author} {\bibfnamefont {R.}~\bibnamefont {Nitsche}}\ and\ \bibinfo {author} {\bibfnamefont {W.~J.}\ \bibnamefont {Merz}},\ }\bibfield  {title} {\bibinfo {title} {\singleletter{Photoconduction in ternary V-VI-VII compounds}},\ }\href {https://doi.org/10.1016/0022-3697(60)90136-0} {\bibfield  {journal} {\bibinfo  {journal} {Journal of Physics and Chemistry of Solids}\ }\textbf {\bibinfo {volume} {13}},\ \bibinfo {pages} {154} (\bibinfo {year} {1960})}\BibitemShut {NoStop}%
\bibitem [{\citenamefont {Fatuzzo}\ \emph {et~al.}(1962)\citenamefont {Fatuzzo}, \citenamefont {Nitsche}, \citenamefont {Harbeke}, \citenamefont {Ruppel}, \citenamefont {Roetschi},\ and\ \citenamefont {Merz}}]{fatuzzo1962a}%
  \BibitemOpen
  \bibfield  {author} {\bibinfo {author} {\bibfnamefont {E.}~\bibnamefont {Fatuzzo}}, \bibinfo {author} {\bibfnamefont {R.}~\bibnamefont {Nitsche}}, \bibinfo {author} {\bibfnamefont {G.}~\bibnamefont {Harbeke}}, \bibinfo {author} {\bibfnamefont {W.}~\bibnamefont {Ruppel}}, \bibinfo {author} {\bibfnamefont {H.}~\bibnamefont {Roetschi}},\ and\ \bibinfo {author} {\bibfnamefont {W.~J.}\ \bibnamefont {Merz}},\ }\bibfield  {title} {\bibinfo {title} {\singleletter{Ferroelectricity in SbSI}},\ }\href {https://doi.org/10.1103/PhysRev.127.2036} {\bibfield  {journal} {\bibinfo  {journal} {Physical Review}\ }\textbf {\bibinfo {volume} {127}},\ \bibinfo {pages} {2036} (\bibinfo {year} {1962})}\BibitemShut {NoStop}%
\bibitem [{\citenamefont {Chen}\ \emph {et~al.}(2015{\natexlab{a}})\citenamefont {Chen}, \citenamefont {Li}, \citenamefont {Yu},\ and\ \citenamefont {Yang}}]{chen2015a}%
  \BibitemOpen
  \bibfield  {author} {\bibinfo {author} {\bibfnamefont {G.}~\bibnamefont {Chen}}, \bibinfo {author} {\bibfnamefont {W.}~\bibnamefont {Li}}, \bibinfo {author} {\bibfnamefont {Y.}~\bibnamefont {Yu}},\ and\ \bibinfo {author} {\bibfnamefont {Q.}~\bibnamefont {Yang}},\ }\bibfield  {title} {\bibinfo {title} {\singleletter{Fast and low-temperature synthesis of one-dimensional (1D) single-crystalline SbSI microrod for high performance photodetector}},\ }\href {https://doi.org/10.1039/c5ra01180a} {\bibfield  {journal} {\bibinfo  {journal} {RSC Advances}\ }\textbf {\bibinfo {volume} {5}},\ \bibinfo {pages} {21859} (\bibinfo {year} {2015}{\natexlab{a}})}\BibitemShut {NoStop}%
\bibitem [{\citenamefont {Gödel}\ and\ \citenamefont {Steiner}(2016)}]{goedel2016a}%
  \BibitemOpen
  \bibfield  {author} {\bibinfo {author} {\bibfnamefont {K.~C.}\ \bibnamefont {Gödel}}\ and\ \bibinfo {author} {\bibfnamefont {U.}~\bibnamefont {Steiner}},\ }\bibfield  {title} {\bibinfo {title} {\singleletter{Thin film synthesis of SbSI micro-crystals for self-powered photodetectors with rapid time response}},\ }\href {https://doi.org/10.1039/c6nr04759a} {\bibfield  {journal} {\bibinfo  {journal} {Nanoscale}\ }\textbf {\bibinfo {volume} {8}},\ \bibinfo {pages} {15920} (\bibinfo {year} {2016})}\BibitemShut {NoStop}%
\bibitem [{\citenamefont {Sun}\ \emph {et~al.}(2019)\citenamefont {Sun}, \citenamefont {Wang}, \citenamefont {Xu}, \citenamefont {Wang}, \citenamefont {Liu}, \citenamefont {Chen},\ and\ \citenamefont {Yi}}]{sun2019a}%
  \BibitemOpen
  \bibfield  {author} {\bibinfo {author} {\bibfnamefont {L.}~\bibnamefont {Sun}}, \bibinfo {author} {\bibfnamefont {C.}~\bibnamefont {Wang}}, \bibinfo {author} {\bibfnamefont {L.}~\bibnamefont {Xu}}, \bibinfo {author} {\bibfnamefont {J.}~\bibnamefont {Wang}}, \bibinfo {author} {\bibfnamefont {X.}~\bibnamefont {Liu}}, \bibinfo {author} {\bibfnamefont {X.}~\bibnamefont {Chen}},\ and\ \bibinfo {author} {\bibfnamefont {G.~C.}\ \bibnamefont {Yi}},\ }\bibfield  {title} {\bibinfo {title} {\singleletter{SbSI whisker/PbI$_\text{2}$ flake mixed-dimensional van der Waals heterostructure for photodetection}},\ }\href {https://doi.org/10.1039/c9ce00544g} {\bibfield  {journal} {\bibinfo  {journal} {CrystEngComm}\ }\textbf {\bibinfo {volume} {21}},\ \bibinfo {pages} {3779} (\bibinfo {year} {2019})}\BibitemShut {NoStop}%
\bibitem [{\citenamefont {Shen}\ \emph {et~al.}(2020)\citenamefont {Shen}, \citenamefont {Liu}, \citenamefont {Wang}, \citenamefont {Wang}, \citenamefont {Wu}, \citenamefont {Chen},\ and\ \citenamefont {Yi}}]{shen2020a}%
  \BibitemOpen
  \bibfield  {author} {\bibinfo {author} {\bibfnamefont {J.}~\bibnamefont {Shen}}, \bibinfo {author} {\bibfnamefont {X.}~\bibnamefont {Liu}}, \bibinfo {author} {\bibfnamefont {C.}~\bibnamefont {Wang}}, \bibinfo {author} {\bibfnamefont {J.}~\bibnamefont {Wang}}, \bibinfo {author} {\bibfnamefont {B.}~\bibnamefont {Wu}}, \bibinfo {author} {\bibfnamefont {X.}~\bibnamefont {Chen}},\ and\ \bibinfo {author} {\bibfnamefont {G.~C.}\ \bibnamefont {Yi}},\ }\bibfield  {title} {\bibinfo {title} {\singleletter{SbSI microrod based flexible photodetectors}},\ }\href {https://doi.org/10.1088/1361-6463/ab8ced} {\bibfield  {journal} {\bibinfo  {journal} {Journal of Physics D: Applied Physics}\ }\textbf {\bibinfo {volume} {53}},\ \bibinfo {pages} {345106} (\bibinfo {year} {2020})}\BibitemShut {NoStop}%
\bibitem [{\citenamefont {Choi}\ \emph {et~al.}(2018)\citenamefont {Choi}, \citenamefont {Hwang},\ and\ \citenamefont {Kim}}]{choi2018a}%
  \BibitemOpen
  \bibfield  {author} {\bibinfo {author} {\bibfnamefont {Y.~C.}\ \bibnamefont {Choi}}, \bibinfo {author} {\bibfnamefont {E.}~\bibnamefont {Hwang}},\ and\ \bibinfo {author} {\bibfnamefont {D.~H.}\ \bibnamefont {Kim}},\ }\bibfield  {title} {\bibinfo {title} {\singleletter{Controlled growth of SbSI thin films from amorphous Sb$_\text{2}$S$_\text{3}$ for low-temperature solution processed chalcohalide solar cells}},\ }\href {https://doi.org/10.1063/1.5058166} {\bibfield  {journal} {\bibinfo  {journal} {APL Materials}\ }\textbf {\bibinfo {volume} {6}},\ \bibinfo {pages} {121108} (\bibinfo {year} {2018})}\BibitemShut {NoStop}%
\bibitem [{\citenamefont {Nie}\ \emph {et~al.}(2018)\citenamefont {Nie}, \citenamefont {Yun}, \citenamefont {Paik}, \citenamefont {Mehta}, \citenamefont {Park}, \citenamefont {Choi},\ and\ \citenamefont {Seok}}]{nie2018a}%
  \BibitemOpen
  \bibfield  {author} {\bibinfo {author} {\bibfnamefont {R.}~\bibnamefont {Nie}}, \bibinfo {author} {\bibfnamefont {H.~S.}\ \bibnamefont {Yun}}, \bibinfo {author} {\bibfnamefont {M.~J.}\ \bibnamefont {Paik}}, \bibinfo {author} {\bibfnamefont {A.}~\bibnamefont {Mehta}}, \bibinfo {author} {\bibfnamefont {B.~W.}\ \bibnamefont {Park}}, \bibinfo {author} {\bibfnamefont {Y.~C.}\ \bibnamefont {Choi}},\ and\ \bibinfo {author} {\bibfnamefont {S.~I.}\ \bibnamefont {Seok}},\ }\bibfield  {title} {\bibinfo {title} {\singleletter{Efficient Solar Cells Based on Light-Harvesting Antimony Sulfoiodide}},\ }\href {https://doi.org/10.1002/aenm.201701901} {\bibfield  {journal} {\bibinfo  {journal} {Advanced Energy Materials}\ }\textbf {\bibinfo {volume} {8}},\ \bibinfo {pages} {1701901} (\bibinfo {year} {2018})}\BibitemShut {NoStop}%
\bibitem [{\citenamefont {Wang}\ \emph {et~al.}(2018)\citenamefont {Wang}, \citenamefont {Zhang}, \citenamefont {Fang}, \citenamefont {Chen}, \citenamefont {Li}, \citenamefont {Sheng}, \citenamefont {Xu}, \citenamefont {Hui}, \citenamefont {Lan}, \citenamefont {Fang}, \citenamefont {Wang}, \citenamefont {Wang}, \citenamefont {Dai}, \citenamefont {Bao},\ and\ \citenamefont {Wang}}]{wang2018a}%
  \BibitemOpen
  \bibfield  {author} {\bibinfo {author} {\bibfnamefont {C.}~\bibnamefont {Wang}}, \bibinfo {author} {\bibfnamefont {M.}~\bibnamefont {Zhang}}, \bibinfo {author} {\bibfnamefont {Y.}~\bibnamefont {Fang}}, \bibinfo {author} {\bibfnamefont {G.}~\bibnamefont {Chen}}, \bibinfo {author} {\bibfnamefont {Q.}~\bibnamefont {Li}}, \bibinfo {author} {\bibfnamefont {X.}~\bibnamefont {Sheng}}, \bibinfo {author} {\bibfnamefont {X.}~\bibnamefont {Xu}}, \bibinfo {author} {\bibfnamefont {J.}~\bibnamefont {Hui}}, \bibinfo {author} {\bibfnamefont {Y.}~\bibnamefont {Lan}}, \bibinfo {author} {\bibfnamefont {M.}~\bibnamefont {Fang}}, \bibinfo {author} {\bibfnamefont {X.}~\bibnamefont {Wang}}, \bibinfo {author} {\bibfnamefont {X.}~\bibnamefont {Wang}}, \bibinfo {author} {\bibfnamefont {Z.}~\bibnamefont {Dai}}, \bibinfo {author} {\bibfnamefont {J.}~\bibnamefont {Bao}},\ and\ \bibinfo {author} {\bibfnamefont {P.}~\bibnamefont {Wang}},\ }\bibfield  {title} {\bibinfo {title} {\singleletter{SbSI Nanocrystals: An Excellent Visible Light
  Photocatalyst with Efficient Generation of Singlet Oxygen}},\ }\href {https://doi.org/10.1021/acssuschemeng.8b02498} {\bibfield  {journal} {\bibinfo  {journal} {ACS Sustainable Chemistry and Engineering}\ }\textbf {\bibinfo {volume} {6}},\ \bibinfo {pages} {12166} (\bibinfo {year} {2018})}\BibitemShut {NoStop}%
\bibitem [{\citenamefont {Tasviri}\ and\ \citenamefont {Sajadi-Hezave}(2017)}]{tasviri2017a}%
  \BibitemOpen
  \bibfield  {author} {\bibinfo {author} {\bibfnamefont {M.}~\bibnamefont {Tasviri}}\ and\ \bibinfo {author} {\bibfnamefont {Z.}~\bibnamefont {Sajadi-Hezave}},\ }\bibfield  {title} {\bibinfo {title} {\singleletter{SbSI nanowires and CNTs encapsulated with SbSI as photocatalysts with high visible-light driven photoactivity}},\ }\href {https://doi.org/10.1016/j.mcat.2017.04.020} {\bibfield  {journal} {\bibinfo  {journal} {Molecular Catalysis}\ }\textbf {\bibinfo {volume} {436}},\ \bibinfo {pages} {174} (\bibinfo {year} {2017})}\BibitemShut {NoStop}%
\bibitem [{\citenamefont {Tamilselvan}\ and\ \citenamefont {Bhattacharyya}(2016)}]{tamilselvan2016a}%
  \BibitemOpen
  \bibfield  {author} {\bibinfo {author} {\bibfnamefont {M.}~\bibnamefont {Tamilselvan}}\ and\ \bibinfo {author} {\bibfnamefont {A.~J.}\ \bibnamefont {Bhattacharyya}},\ }\bibfield  {title} {\bibinfo {title} {\singleletter{Antimony sulphoiodide (SbSI), a narrow band-gap non-oxide ternary semiconductor with efficient photocatalytic activity}},\ }\href {https://doi.org/10.1039/c6ra23750a} {\bibfield  {journal} {\bibinfo  {journal} {RSC Advances}\ }\textbf {\bibinfo {volume} {6}},\ \bibinfo {pages} {105980} (\bibinfo {year} {2016})}\BibitemShut {NoStop}%
\bibitem [{\citenamefont {Kwolek}\ \emph {et~al.}(2015)\citenamefont {Kwolek}, \citenamefont {Pilarczyk}, \citenamefont {Tokarski}, \citenamefont {Mech}, \citenamefont {Irzmański},\ and\ \citenamefont {Szaciłowski}}]{kwolek2015a}%
  \BibitemOpen
  \bibfield  {author} {\bibinfo {author} {\bibfnamefont {P.}~\bibnamefont {Kwolek}}, \bibinfo {author} {\bibfnamefont {K.}~\bibnamefont {Pilarczyk}}, \bibinfo {author} {\bibfnamefont {T.}~\bibnamefont {Tokarski}}, \bibinfo {author} {\bibfnamefont {J.}~\bibnamefont {Mech}}, \bibinfo {author} {\bibfnamefont {J.}~\bibnamefont {Irzmański}},\ and\ \bibinfo {author} {\bibfnamefont {K.}~\bibnamefont {Szaciłowski}},\ }\bibfield  {title} {\bibinfo {title} {\singleletter{Photoelectrochemistry of n-type antimony sulfoiodide nanowires}},\ }\href {https://doi.org/10.1088/0957-4484/26/10/105710} {\bibfield  {journal} {\bibinfo  {journal} {Nanotechnology}\ }\textbf {\bibinfo {volume} {26}},\ \bibinfo {pages} {105710} (\bibinfo {year} {2015})}\BibitemShut {NoStop}%
\bibitem [{\citenamefont {Purusothaman}\ \emph {et~al.}(2018)\citenamefont {Purusothaman}, \citenamefont {Alluri}, \citenamefont {Chandrasekhar},\ and\ \citenamefont {Kim}}]{purusothaman2018a}%
  \BibitemOpen
  \bibfield  {author} {\bibinfo {author} {\bibfnamefont {Y.}~\bibnamefont {Purusothaman}}, \bibinfo {author} {\bibfnamefont {N.~R.}\ \bibnamefont {Alluri}}, \bibinfo {author} {\bibfnamefont {A.}~\bibnamefont {Chandrasekhar}},\ and\ \bibinfo {author} {\bibfnamefont {S.~J.}\ \bibnamefont {Kim}},\ }\bibfield  {title} {\bibinfo {title} {\singleletter{Photoactive piezoelectric energy harvester driven by antimony sulfoiodide (SbSI): A A$_\text{V}$B$_\text{VI}$C$_\text{VII}$ class ferroelectric-semiconductor compound}},\ }\href {https://doi.org/10.1016/j.nanoen.2018.05.058} {\bibfield  {journal} {\bibinfo  {journal} {Nano Energy}\ }\textbf {\bibinfo {volume} {50}},\ \bibinfo {pages} {256} (\bibinfo {year} {2018})}\BibitemShut {NoStop}%
\bibitem [{\citenamefont {Chen}\ \emph {et~al.}(2015{\natexlab{b}})\citenamefont {Chen}, \citenamefont {Li}, \citenamefont {Zhou}, \citenamefont {Chen}, \citenamefont {Luo}, \citenamefont {Liu}, \citenamefont {Zeng}, \citenamefont {Yang}, \citenamefont {Zhang}, \citenamefont {Han},\ and\ \citenamefont {Tang}}]{chen2015b}%
  \BibitemOpen
  \bibfield  {author} {\bibinfo {author} {\bibfnamefont {C.}~\bibnamefont {Chen}}, \bibinfo {author} {\bibfnamefont {W.}~\bibnamefont {Li}}, \bibinfo {author} {\bibfnamefont {Y.}~\bibnamefont {Zhou}}, \bibinfo {author} {\bibfnamefont {C.}~\bibnamefont {Chen}}, \bibinfo {author} {\bibfnamefont {M.}~\bibnamefont {Luo}}, \bibinfo {author} {\bibfnamefont {X.}~\bibnamefont {Liu}}, \bibinfo {author} {\bibfnamefont {K.}~\bibnamefont {Zeng}}, \bibinfo {author} {\bibfnamefont {B.}~\bibnamefont {Yang}}, \bibinfo {author} {\bibfnamefont {C.}~\bibnamefont {Zhang}}, \bibinfo {author} {\bibfnamefont {J.}~\bibnamefont {Han}},\ and\ \bibinfo {author} {\bibfnamefont {J.}~\bibnamefont {Tang}},\ }\bibfield  {title} {\bibinfo {title} {\singleletter{Optical properties of amorphous and polycrystalline Sb$_\text{2}$Se$_\text{3}$ thin films prepared by thermal evaporation}},\ }\href {https://doi.org/10.1063/1.4927741} {\bibfield  {journal} {\bibinfo  {journal} {Applied Physics Letters}\ }\textbf {\bibinfo {volume} {107}},\ \bibinfo
  {pages} {043905} (\bibinfo {year} {2015}{\natexlab{b}})}\BibitemShut {NoStop}%
\bibitem [{\citenamefont {Versavel}\ and\ \citenamefont {Haber}(2007)}]{versavel2007a}%
  \BibitemOpen
  \bibfield  {author} {\bibinfo {author} {\bibfnamefont {M.~Y.}\ \bibnamefont {Versavel}}\ and\ \bibinfo {author} {\bibfnamefont {J.~A.}\ \bibnamefont {Haber}},\ }\bibfield  {title} {\bibinfo {title} {\singleletter{Structural and optical properties of amorphous and crystalline antimony sulfide thin-films}},\ }\href {https://doi.org/10.1016/j.tsf.2007.03.043} {\bibfield  {journal} {\bibinfo  {journal} {Thin Solid Films}\ }\textbf {\bibinfo {volume} {515}},\ \bibinfo {pages} {7171} (\bibinfo {year} {2007})}\BibitemShut {NoStop}%
\bibitem [{\citenamefont {Levanyuk}\ and\ \citenamefont {Osipov}(1981)}]{levanyuk1981a}%
  \BibitemOpen
  \bibfield  {author} {\bibinfo {author} {\bibfnamefont {A.~P.}\ \bibnamefont {Levanyuk}}\ and\ \bibinfo {author} {\bibfnamefont {V.~V.}\ \bibnamefont {Osipov}},\ }\bibfield  {title} {\bibinfo {title} {\singleletter{Edge luminescence of direct-gap semiconductors}},\ }\href {https://doi.org/10.1070/PU1981v024n03ABEH004770} {\bibfield  {journal} {\bibinfo  {journal} {Soviet Physics Uspekhi}\ }\textbf {\bibinfo {volume} {24}},\ \bibinfo {pages} {187} (\bibinfo {year} {1981})}\BibitemShut {NoStop}%
\bibitem [{\citenamefont {Schmidt}\ \emph {et~al.}(1992)\citenamefont {Schmidt}, \citenamefont {Lischka},\ and\ \citenamefont {Zulehner}}]{schmidt1992a}%
  \BibitemOpen
  \bibfield  {author} {\bibinfo {author} {\bibfnamefont {T.}~\bibnamefont {Schmidt}}, \bibinfo {author} {\bibfnamefont {K.}~\bibnamefont {Lischka}},\ and\ \bibinfo {author} {\bibfnamefont {W.}~\bibnamefont {Zulehner}},\ }\bibfield  {title} {\bibinfo {title} {\singleletter{Excitation-power dependence of the near-band-edge photoluminescence of semiconductors}},\ }\href {https://doi.org/10.1103/physrevb.45.8989} {\bibfield  {journal} {\bibinfo  {journal} {Physical Review B}\ }\textbf {\bibinfo {volume} {45}},\ \bibinfo {pages} {8989} (\bibinfo {year} {1992})}\BibitemShut {NoStop}%
\bibitem [{\citenamefont {López}\ \emph {et~al.}(2025)\citenamefont {López}, \citenamefont {Kavanagh}, \citenamefont {Benítez}, \citenamefont {Saucedo}, \citenamefont {Walsh}, \citenamefont {Scanlon},\ and\ \citenamefont {Cazorla}}]{lópez2025chalcogenvacanciesrulecharge}%
  \BibitemOpen
  \bibfield  {author} {\bibinfo {author} {\bibfnamefont {C.}~\bibnamefont {López}}, \bibinfo {author} {\bibfnamefont {S.~R.}\ \bibnamefont {Kavanagh}}, \bibinfo {author} {\bibfnamefont {P.}~\bibnamefont {Benítez}}, \bibinfo {author} {\bibfnamefont {E.}~\bibnamefont {Saucedo}}, \bibinfo {author} {\bibfnamefont {A.}~\bibnamefont {Walsh}}, \bibinfo {author} {\bibfnamefont {D.~O.}\ \bibnamefont {Scanlon}},\ and\ \bibinfo {author} {\bibfnamefont {C.}~\bibnamefont {Cazorla}},\ }\bibfield  {title} {\bibinfo {title} {\singleletter{Chalcogen Vacancies Rule Charge Recombination in Pnictogen Chalcohalide Solar-Cell Absorbers}},\ }\href {https://doi.org/10.1021/acsenergylett.5c01267} {\bibfield  {journal} {\bibinfo  {journal} {ACS Energy Letters}\ }\textbf {\bibinfo {volume} {10}},\ \bibinfo {pages} {3562} (\bibinfo {year} {2025})}\BibitemShut {NoStop}%
\bibitem [{\citenamefont {Zhang}\ \emph {et~al.}(2022)\citenamefont {Zhang}, \citenamefont {Wang}, \citenamefont {Gao},\ and\ \citenamefont {Zhao}}]{zhang2022a}%
  \BibitemOpen
  \bibfield  {author} {\bibinfo {author} {\bibfnamefont {F.}~\bibnamefont {Zhang}}, \bibinfo {author} {\bibfnamefont {X.}~\bibnamefont {Wang}}, \bibinfo {author} {\bibfnamefont {W.}~\bibnamefont {Gao}},\ and\ \bibinfo {author} {\bibfnamefont {J.}~\bibnamefont {Zhao}},\ }\bibfield  {title} {\bibinfo {title} {\singleletter{Phonon-Assisted Nonradiative Recombination Tuned by Organic Cations in Ruddlesden-Popper Hybrid Perovskites}},\ }\href {https://doi.org/10.1103/PhysRevApplied.17.064016} {\bibfield  {journal} {\bibinfo  {journal} {Physical Review Applied}\ }\textbf {\bibinfo {volume} {17}},\ \bibinfo {pages} {064016} (\bibinfo {year} {2022})}\BibitemShut {NoStop}%
\bibitem [{\citenamefont {Monserrat}\ \emph {et~al.}(2018)\citenamefont {Monserrat}, \citenamefont {Park}, \citenamefont {Kim},\ and\ \citenamefont {Walsh}}]{monserrat2018a}%
  \BibitemOpen
  \bibfield  {author} {\bibinfo {author} {\bibfnamefont {B.}~\bibnamefont {Monserrat}}, \bibinfo {author} {\bibfnamefont {J.~S.}\ \bibnamefont {Park}}, \bibinfo {author} {\bibfnamefont {S.}~\bibnamefont {Kim}},\ and\ \bibinfo {author} {\bibfnamefont {A.}~\bibnamefont {Walsh}},\ }\bibfield  {title} {\bibinfo {title} {\singleletter{Role of electron-phonon coupling and thermal expansion on band gaps, carrier mobility, and interfacial offsets in kesterite thin-film solar cells}},\ }\href {https://doi.org/10.1063/1.5028186} {\bibfield  {journal} {\bibinfo  {journal} {Applied Physics Letters}\ }\textbf {\bibinfo {volume} {112}},\ \bibinfo {pages} {193903} (\bibinfo {year} {2018})}\BibitemShut {NoStop}%
\bibitem [{\citenamefont {Robert}\ \emph {et~al.}(2020)\citenamefont {Robert}, \citenamefont {Han}, \citenamefont {Kapuscinski}, \citenamefont {Delhomme}, \citenamefont {Faugeras}, \citenamefont {Amand}, \citenamefont {Molas}, \citenamefont {Bartos}, \citenamefont {Watanabe}, \citenamefont {Taniguchi}, \citenamefont {Urbaszek}, \citenamefont {Potemski},\ and\ \citenamefont {Marie}}]{robert2020a}%
  \BibitemOpen
  \bibfield  {author} {\bibinfo {author} {\bibfnamefont {C.}~\bibnamefont {Robert}}, \bibinfo {author} {\bibfnamefont {B.}~\bibnamefont {Han}}, \bibinfo {author} {\bibfnamefont {P.}~\bibnamefont {Kapuscinski}}, \bibinfo {author} {\bibfnamefont {A.}~\bibnamefont {Delhomme}}, \bibinfo {author} {\bibfnamefont {C.}~\bibnamefont {Faugeras}}, \bibinfo {author} {\bibfnamefont {T.}~\bibnamefont {Amand}}, \bibinfo {author} {\bibfnamefont {M.~R.}\ \bibnamefont {Molas}}, \bibinfo {author} {\bibfnamefont {M.}~\bibnamefont {Bartos}}, \bibinfo {author} {\bibfnamefont {K.}~\bibnamefont {Watanabe}}, \bibinfo {author} {\bibfnamefont {T.}~\bibnamefont {Taniguchi}}, \bibinfo {author} {\bibfnamefont {B.}~\bibnamefont {Urbaszek}}, \bibinfo {author} {\bibfnamefont {M.}~\bibnamefont {Potemski}},\ and\ \bibinfo {author} {\bibfnamefont {X.}~\bibnamefont {Marie}},\ }\bibfield  {title} {\bibinfo {title} {\singleletter{Measurement of the spin-forbidden dark excitons in MoS$_\text{2}$ and MoSe$_\text{2}$ monolayers}},\ }\href
  {https://doi.org/10.1038/s41467-020-17608-4} {\bibfield  {journal} {\bibinfo  {journal} {Nature Communications}\ }\textbf {\bibinfo {volume} {11}},\ \bibinfo {pages} {4037} (\bibinfo {year} {2020})}\BibitemShut {NoStop}%
\bibitem [{\citenamefont {Nguyen}\ \emph {et~al.}(2024)\citenamefont {Nguyen}, \citenamefont {Hammel}, \citenamefont {Sharp}, \citenamefont {Kline}, \citenamefont {Schwartz}, \citenamefont {Harvey}, \citenamefont {Nishiwaki}, \citenamefont {Sandeno}, \citenamefont {Ginger}, \citenamefont {Majumdar}, \citenamefont {Yazdi}, \citenamefont {Dukovic},\ and\ \citenamefont {Cossairt}}]{nguyen2024a}%
  \BibitemOpen
  \bibfield  {author} {\bibinfo {author} {\bibfnamefont {H.~A.}\ \bibnamefont {Nguyen}}, \bibinfo {author} {\bibfnamefont {B.~F.}\ \bibnamefont {Hammel}}, \bibinfo {author} {\bibfnamefont {D.}~\bibnamefont {Sharp}}, \bibinfo {author} {\bibfnamefont {J.}~\bibnamefont {Kline}}, \bibinfo {author} {\bibfnamefont {G.}~\bibnamefont {Schwartz}}, \bibinfo {author} {\bibfnamefont {S.}~\bibnamefont {Harvey}}, \bibinfo {author} {\bibfnamefont {E.}~\bibnamefont {Nishiwaki}}, \bibinfo {author} {\bibfnamefont {S.~F.}\ \bibnamefont {Sandeno}}, \bibinfo {author} {\bibfnamefont {D.~S.}\ \bibnamefont {Ginger}}, \bibinfo {author} {\bibfnamefont {A.}~\bibnamefont {Majumdar}}, \bibinfo {author} {\bibfnamefont {S.}~\bibnamefont {Yazdi}}, \bibinfo {author} {\bibfnamefont {G.}~\bibnamefont {Dukovic}},\ and\ \bibinfo {author} {\bibfnamefont {B.~M.}\ \bibnamefont {Cossairt}},\ }\bibfield  {title} {\bibinfo {title} {\singleletter{Colossal Core/Shell CdSe/CdS Quantum Dot Emitters}},\ }\href {https://doi.org/10.1021/acsnano.4c06961}
  {\bibfield  {journal} {\bibinfo  {journal} {ACS Nano}\ }\textbf {\bibinfo {volume} {18}},\ \bibinfo {pages} {20726} (\bibinfo {year} {2024})}\BibitemShut {NoStop}%
\bibitem [{\citenamefont {Liu}\ \emph {et~al.}(2014)\citenamefont {Liu}, \citenamefont {Smith}, \citenamefont {Athanasiou}, \citenamefont {Yu}, \citenamefont {Bai},\ and\ \citenamefont {Wang}}]{liu2014a}%
  \BibitemOpen
  \bibfield  {author} {\bibinfo {author} {\bibfnamefont {B.}~\bibnamefont {Liu}}, \bibinfo {author} {\bibfnamefont {R.}~\bibnamefont {Smith}}, \bibinfo {author} {\bibfnamefont {M.}~\bibnamefont {Athanasiou}}, \bibinfo {author} {\bibfnamefont {X.}~\bibnamefont {Yu}}, \bibinfo {author} {\bibfnamefont {J.}~\bibnamefont {Bai}},\ and\ \bibinfo {author} {\bibfnamefont {T.}~\bibnamefont {Wang}},\ }\bibfield  {title} {\bibinfo {title} {\singleletter{Temporally and spatially resolved photoluminescence investigation of (11$\bar{2}$2) semi-polar InGaN/GaN multiple quantum wells grown on nanorod templates}},\ }\href {https://doi.org/10.1063/1.4905191} {\bibfield  {journal} {\bibinfo  {journal} {Applied Physics Letters}\ }\textbf {\bibinfo {volume} {105}},\ \bibinfo {pages} {261103} (\bibinfo {year} {2014})}\BibitemShut {NoStop}%
\bibitem [{\citenamefont {Yuan}\ \emph {et~al.}(2024)\citenamefont {Yuan}, \citenamefont {Yan}, \citenamefont {Dreessen}, \citenamefont {Rudolph}, \citenamefont {Hülsbeck}, \citenamefont {Klingebiel}, \citenamefont {Ye}, \citenamefont {Rau},\ and\ \citenamefont {Kirchartz}}]{yuan2024a}%
  \BibitemOpen
  \bibfield  {author} {\bibinfo {author} {\bibfnamefont {Y.}~\bibnamefont {Yuan}}, \bibinfo {author} {\bibfnamefont {G.}~\bibnamefont {Yan}}, \bibinfo {author} {\bibfnamefont {C.}~\bibnamefont {Dreessen}}, \bibinfo {author} {\bibfnamefont {T.}~\bibnamefont {Rudolph}}, \bibinfo {author} {\bibfnamefont {M.}~\bibnamefont {Hülsbeck}}, \bibinfo {author} {\bibfnamefont {B.}~\bibnamefont {Klingebiel}}, \bibinfo {author} {\bibfnamefont {J.}~\bibnamefont {Ye}}, \bibinfo {author} {\bibfnamefont {U.}~\bibnamefont {Rau}},\ and\ \bibinfo {author} {\bibfnamefont {T.}~\bibnamefont {Kirchartz}},\ }\bibfield  {title} {\bibinfo {title} {\singleletter{Shallow defects and variable photoluminescence decay times up to 280 µs in triple-cation perovskites}},\ }\href {https://doi.org/10.1038/s41563-023-01771-2} {\bibfield  {journal} {\bibinfo  {journal} {Nature Materials}\ }\textbf {\bibinfo {volume} {23}},\ \bibinfo {pages} {391} (\bibinfo {year} {2024})}\BibitemShut {NoStop}%
\end{thebibliography}%

\end{document}

%\preprint{APS/123-QED}

\title{\Large{SUPPORTING INFORMATION} \\ \vspace{1cm} \large Parallel Exploration of the Optoelectronic Properties of (Sb,Bi)(S,Se)(Br,I) Chalcohalides}% Force line breaks with \\

\author{Rasmus S. Nielsen}
\email[]{Electronic mail: rasmus.nielsen@empa.ch}
\affiliation{Nanomaterials Spectroscopy and Imaging, Transport at Nanoscale Interfaces Laboratory, Swiss Federal Laboratories for Material Science and Technology (EMPA), Ueberlandstrasse 129, 8600 Duebendorf, Switzerland}

\author{Ángel Labordet Álvarez}
\affiliation{Nanomaterials Spectroscopy and Imaging, Transport at Nanoscale Interfaces Laboratory, Swiss Federal Laboratories for Material Science and Technology (EMPA), Ueberlandstrasse 129, 8600 Duebendorf, Switzerland}
\affiliation{Department of Physics, University of Basel, 4056 Basel, Switzerland}
\affiliation{Swiss Nanoscience Institute, University of Basel, 4056 Basel, Switzerland}

\author{Axel G. Medaille}
\affiliation{Universitat Politècnica de Catalunya (UPC), Photovoltaic Lab - Micro and Nano Technologies Group (MNT), Electronic Engineering Department, EEBE, Av Eduard Maristany 10-14, Barcelona 08019, Spain}
\affiliation{Universitat Politècnica de Catalunya (UPC), Barcelona Centre for Multiscale Science \& Engineering, Av Eduard Maristany 10-14, Barcelona 08019, Spain}

\author{Ivan Caño}
\affiliation{Universitat Politècnica de Catalunya (UPC), Photovoltaic Lab - Micro and Nano Technologies Group (MNT), Electronic Engineering Department, EEBE, Av Eduard Maristany 10-14, Barcelona 08019, Spain}
\affiliation{Universitat Politècnica de Catalunya (UPC), Barcelona Centre for Multiscale Science \& Engineering, Av Eduard Maristany 10-14, Barcelona 08019, Spain}

\author{Alejandro Navarro-Güell}
\affiliation{Universitat Politècnica de Catalunya (UPC), Photovoltaic Lab - Micro and Nano Technologies Group (MNT), Electronic Engineering Department, EEBE, Av Eduard Maristany 10-14, Barcelona 08019, Spain}
\affiliation{Universitat Politècnica de Catalunya (UPC), Barcelona Centre for Multiscale Science \& Engineering, Av Eduard Maristany 10-14, Barcelona 08019, Spain}

\author{Cibrán L. Álvarez}
\affiliation{Group of Characterization of Materials, Departament de Física, Universitat Politècnica de Catalunya (UPC), Campus Diagonal-Besòs, Av Eduard Maristany 10–14, Barcelona 08019, Spain}
\affiliation{Universitat Politècnica de Catalunya (UPC), Barcelona Centre for Multiscale Science \& Engineering, Av Eduard Maristany 10-14, Barcelona 08019, Spain}

\author{Claudio Cazorla}
\affiliation{Group of Characterization of Materials, Departament de Física, Universitat Politècnica de Catalunya (UPC), Campus Diagonal-Besòs, Av Eduard Maristany 10–14, Barcelona 08019, Spain}
\affiliation{Universitat Politècnica de Catalunya (UPC), Barcelona Centre for Multiscale Science \& Engineering, Av Eduard Maristany 10-14, Barcelona 08019, Spain}

\author{David R. Ferrer}
\affiliation{Universitat Politècnica de Catalunya (UPC), Photovoltaic Lab - Micro and Nano Technologies Group (MNT), Electronic Engineering Department, EEBE, Av Eduard Maristany 10-14, Barcelona 08019, Spain}
\affiliation{Universitat Politècnica de Catalunya (UPC), Barcelona Centre for Multiscale Science \& Engineering, Av Eduard Maristany 10-14, Barcelona 08019, Spain}

\author{Zacharie J. Li-Kao}
\affiliation{Universitat Politècnica de Catalunya (UPC), Photovoltaic Lab - Micro and Nano Technologies Group (MNT), Electronic Engineering Department, EEBE, Av Eduard Maristany 10-14, Barcelona 08019, Spain}
\affiliation{Universitat Politècnica de Catalunya (UPC), Barcelona Centre for Multiscale Science \& Engineering, Av Eduard Maristany 10-14, Barcelona 08019, Spain}

\author{Edgardo Saucedo}
\affiliation{Universitat Politècnica de Catalunya (UPC), Photovoltaic Lab - Micro and Nano Technologies Group (MNT), Electronic Engineering Department, EEBE, Av Eduard Maristany 10-14, Barcelona 08019, Spain}
\affiliation{Universitat Politècnica de Catalunya (UPC), Barcelona Centre for Multiscale Science \& Engineering, Av Eduard Maristany 10-14, Barcelona 08019, Spain}

\author{Mirjana Dimitrievska}
\email[]{Electronic mail: mirjana.dimitrievska@empa.ch}
\affiliation{Nanomaterials Spectroscopy and Imaging, Transport at Nanoscale Interfaces Laboratory, Swiss Federal Laboratories for Material Science and Technology (EMPA), Ueberlandstrasse 129, 8600 Duebendorf, Switzerland}

\maketitle

\vfill

\clearpage

\begin{figure*}[t!]
    \centering
    \includegraphics[width=\textwidth,trim={20 45 20 125},clip]{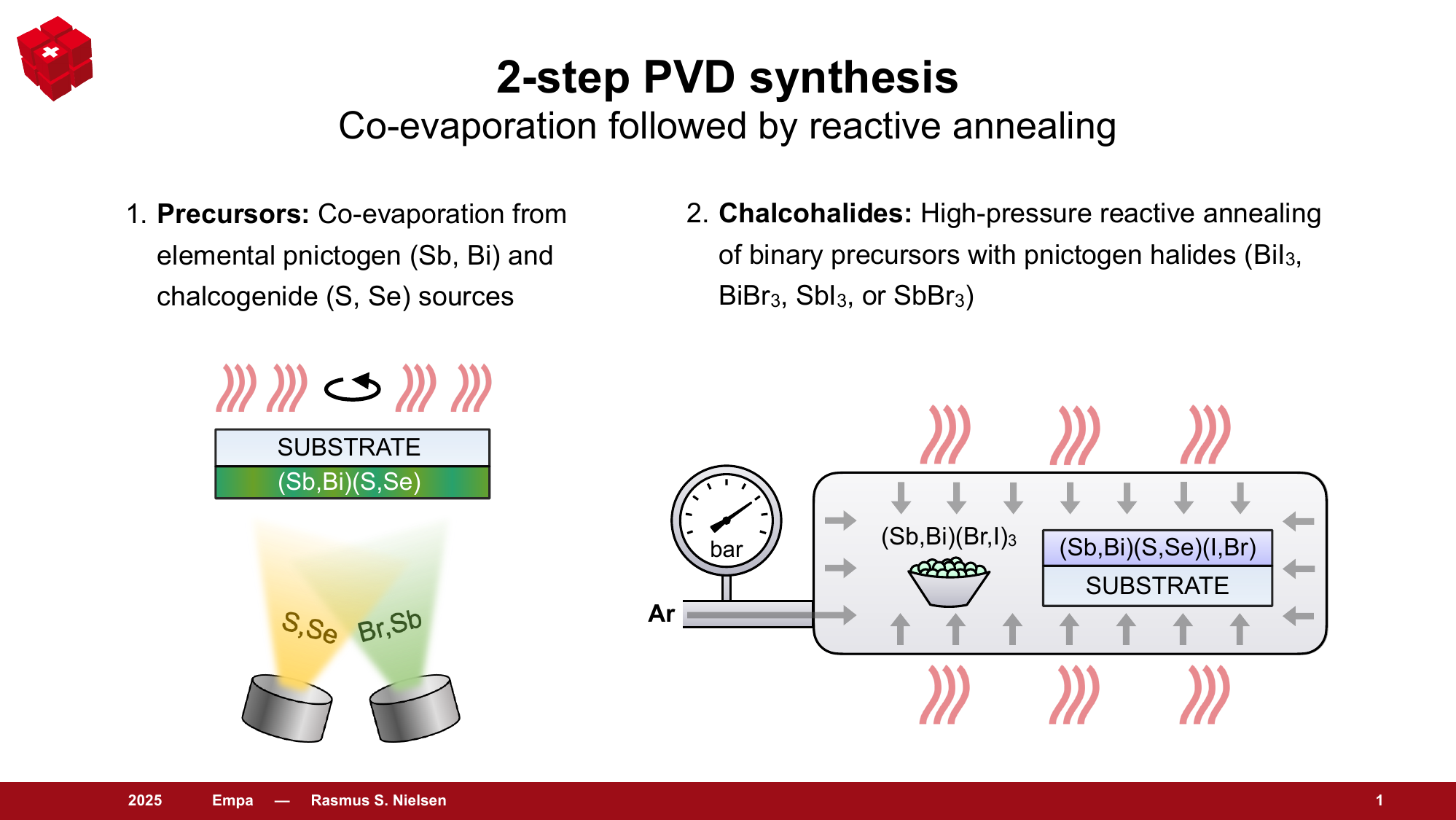}%TRIM 0 0 0 0
    \caption{Schematic illustration of the two-step physical vapor deposition (PVD) process used to synthesize pnictogen chalcohalides. In the first step, a binary precursor thin film is deposited via thermal co-evaporation of elemental pnictogen and chalcogenide sources. In the second step, the precursor film is placed in a pressurized furnace with pnictogen halide source materials, where high-pressure reactive annealing converts the film into the desired ternary chalcohalide phase.}
    \label{fig:ESI_Figure4}
\end{figure*}

\clearpage

\begin{figure*}[t!]
    \centering
    \includegraphics[width=\textwidth,trim={0 0 0 0},clip]{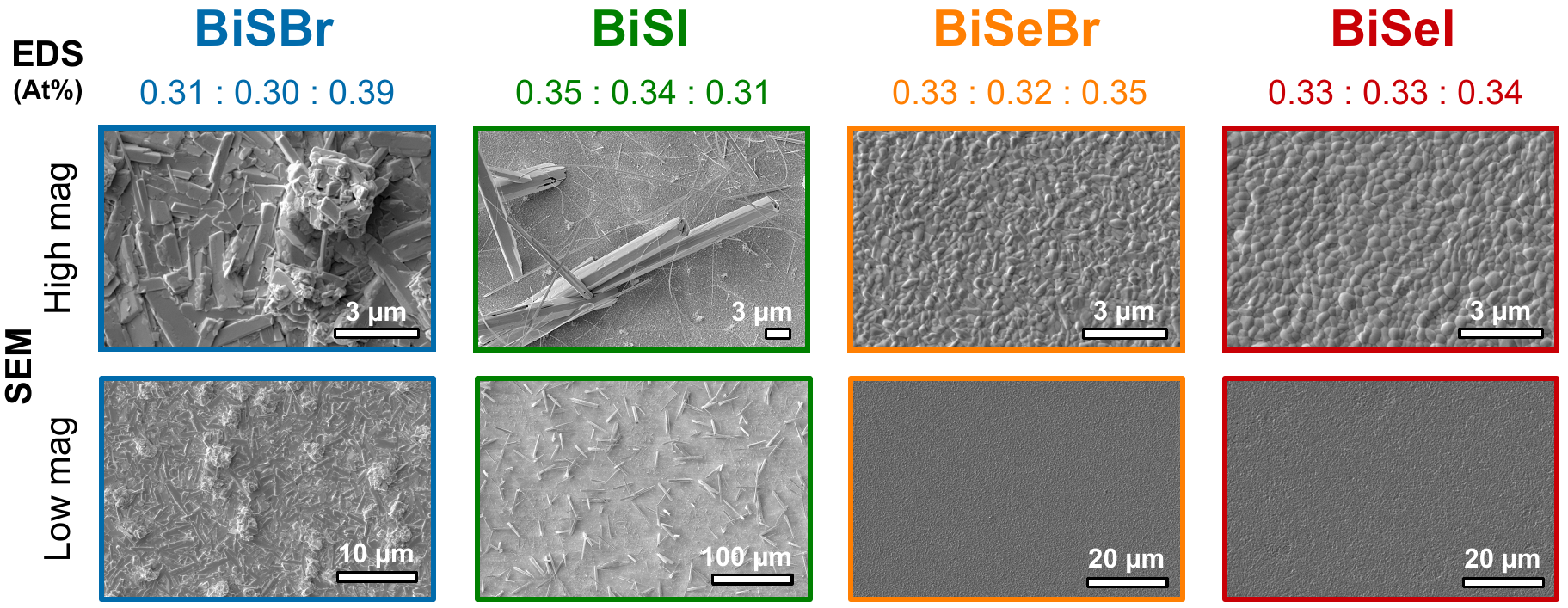}\vspace{1cm} \newpage%TRIM 35 58 49 143
    \includegraphics[width=\textwidth,trim={0 0 0 0},clip]{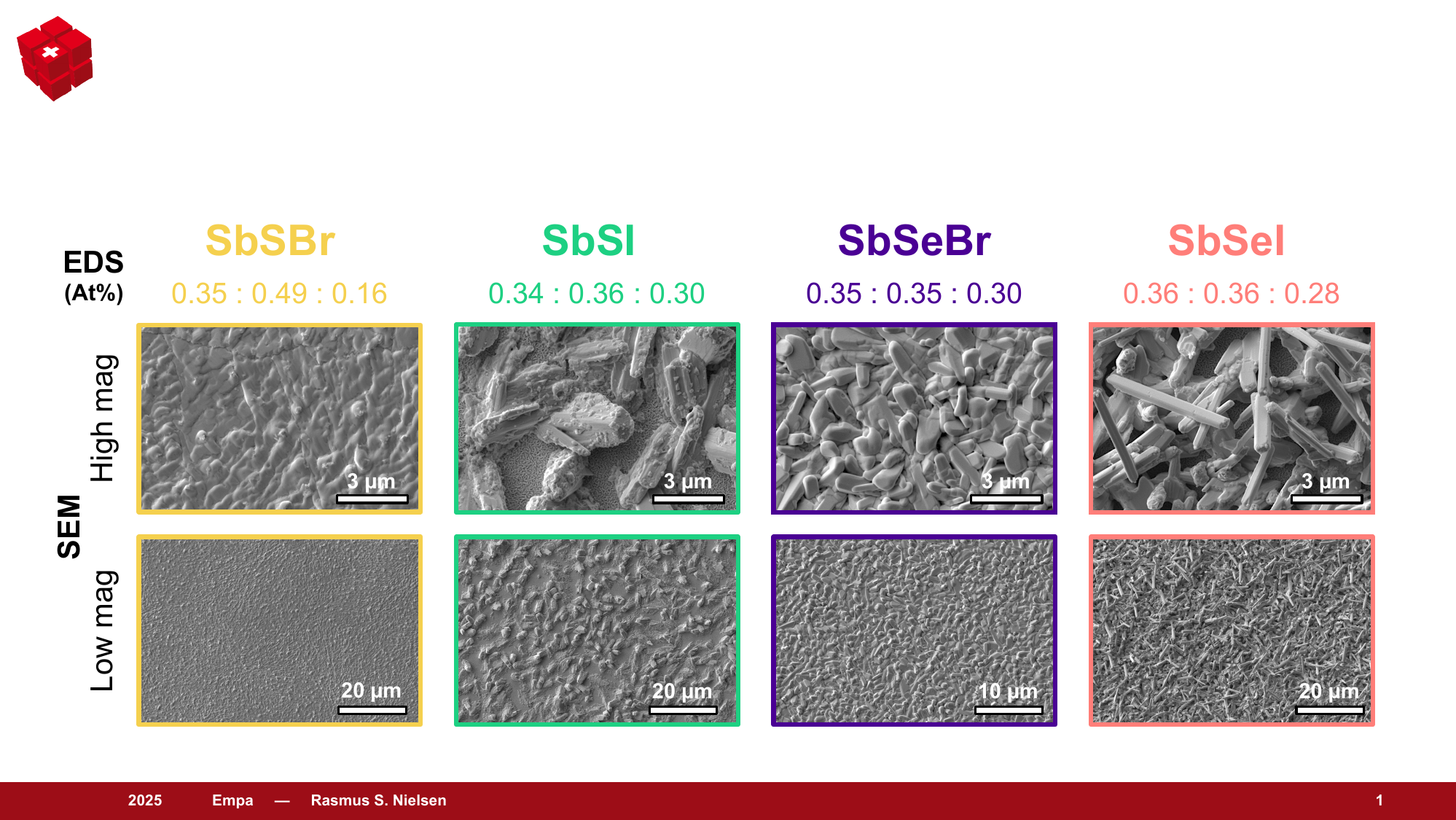}%TRIM 35 58 49 143
    \caption{Top view SEM images at high and low magnification for all Bi- and Sb-based chalcohalides, along with corresponding atomic percentages from EDS analysis.}
    \label{fig:ESI_Figure1}
\end{figure*}

\clearpage

\begin{figure*}[t!]
    \centering
    \includegraphics[width=\textwidth,trim={0 0 0 0},clip]{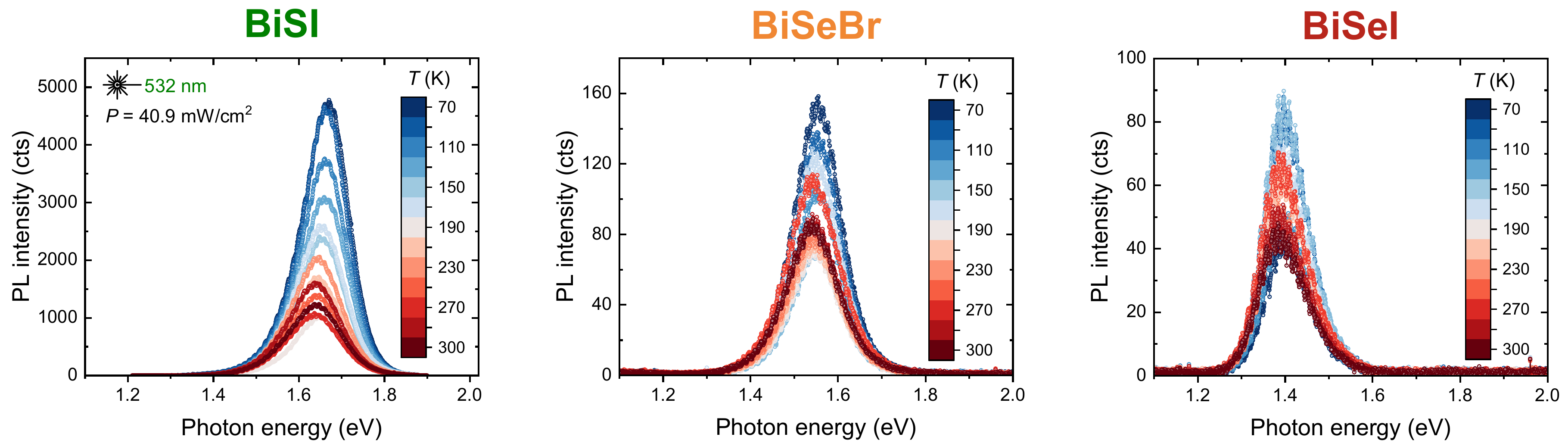}%TRIM 0 0 0 0
    \caption{Temperature-dependent PL spectra of BiSI, BiSeI, and BiSeBr from 70 to 300 K under 532 nm excitation at 40.9 mW/cm$^\text{2}$.}
    \label{fig:ESI_Figure2}
\end{figure*}

\clearpage

\begin{figure*}[t!]
    \centering
    \includegraphics[width=\textwidth,trim={0 0 0 0},clip]{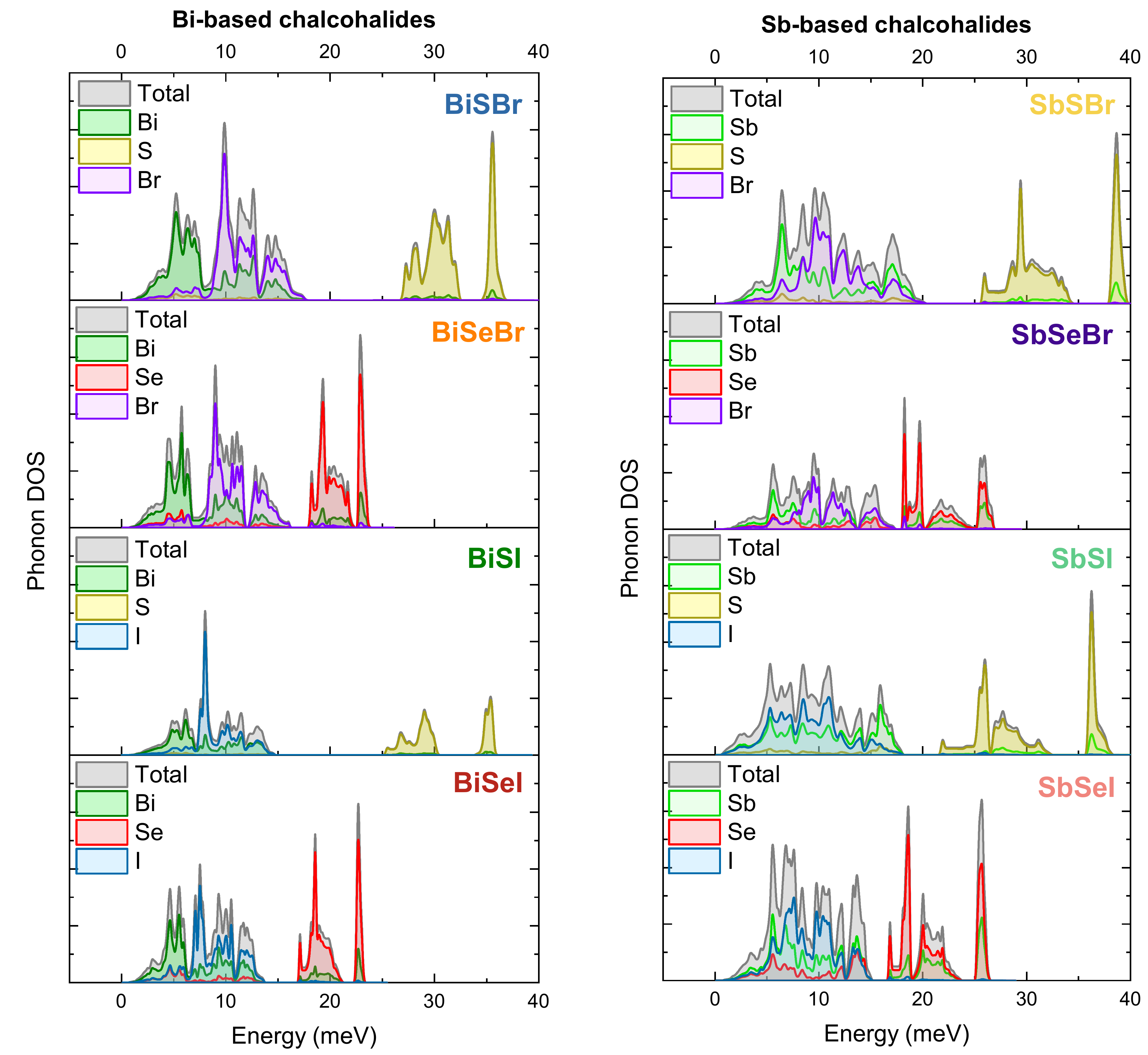}%TRIM 0 0 0 0
    \caption{Projected phonon density of states (DOS) for all Bi- and Sb-based chalcohalides. All sulfur-based compounds exhibit a phonon gap below the high-frequency chalcogen band, while selenium-based compounds show Se-related modes that fill this gap, enhancing phonon scattering pathways.}
    \label{fig:ESI_Figure3}
\end{figure*}

\clearpage

\begin{table}[ht!]
\small
\caption{Room-temperature bandgap energies for all eight chalcohalide compounds, determined from the photoluminescence peak positions. The value for SbSeBr is marked with an asterisk (*) to indicate that the structural determination for this compound was inconclusive; therefore, the observed deviation from the sulfur-selenium trend should be interpreted with caution.}
\vspace{0.25cm}
\label{tbl:SupplementaryPLPeakEnergies}
\begin{tabular*}{0.25\columnwidth}{@{\extracolsep{\fill}}lc} % Adjusted column alignment
    Compound & $E_\mathrm{g}$ (eV) \\ \hline \vspace{-0.25cm} \\
    BiSeI & 1.38 \\
    BiSeBr & 1.54 \\
    BiSI & 1.64 \\
    BiSBr\vspace{0.1cm} & 1.99 \\
    SbSeI & 1.74 \\
    SbSBr & 1.81 \\ 
    SbSI & 1.91 \\
    SbSeBr* & 2.08 \\
    \hline 
\end{tabular*}
\end{table}

\clearpage

%\nocite{*}

%\bibliography{references}% Produces the bibliography via BibTeX.